\magnification=\magstep1
\baselineskip 12 pt
\parskip 6pt
\overfullrule = 0pt
\def\cl{\centerline}
\def\np{\vfill\eject}


\def\kms{\ {\rm km\,s^{-1}}}

\def\dd{{\rm d}}
\def\nb{{\bar n}}
\def\ln{{\rm ln}}

\def\pmb#1{\setbox0=\hbox{#1}%
\kern-.025em\copy0\kern-\wd0
\kern.05em\copy0\kern-\wd0
\kern-.025em\raise.0433em\box0}

\def\vv{\pmb{$v$}}
\def\vs{\pmb{$s$}}

\def\vaj{\pmb{$\alpha$}}

\def\iras {{\it IRAS~}}
\def\velmod{VELMOD~}
\def\Ft {{\tilde F}}
\def\tj {{\tilde F}}
\def\aj {{\alpha}^j}
\def\ajp {{\alpha}^{j'}}
\def\ip {i^\prime}
\def\jp {j^\prime}
\def\bfr {{\bf r}}
\def\etal{{\it et al.\ }}

\def\simlt{\lower.5ex\hbox{$\; \buildrel < \over \sim \;$}}
\def\simgt{\lower.5ex\hbox{$\; \buildrel > \over \sim \;$}}
\def\Msun{\rm\> M_{\odot}}
\font\bigbf=cmbx12 at 14.4truept
\def\vnabla{\pmb{$\nabla$}}
\def\myfig#1#2#3#4{
  \midinsert
     \vskip #2 true cm \vskip 1.3 true cm
     #3
     \vskip -2. true cm
     {\baselineskip 11pt
       \rightskip=0.8 true cm \leftskip=0.8 true cm
       \par\noindent{\bf Figure #1:}  #4 \par
       \rightskip=0 true cm \leftskip=0 true cm
     }
     \vskip -.1 true cm
  \endinsert
}    
       
%

%
\line{\hfil {\it submitted to the Astrophysical Journal, April 1996}}
\line{\hfil {\it accepted, May 1996}}

{\cl {\bigbf COMPARISON OF VELOCITY AND GRAVITY FIELDS:}}
{\cl {\bigbf THE MARK III TULLY-FISHER CATALOG }}
{\cl {{\bigbf VERSUS THE IRAS 1.2 JY SURVEY}\footnote *{
The manuscript with all its embedded figures (many in color) 
can be retrieved from \break
{\it ftp://magicbean.berkeley.edu/pub/marc/m3itf/vitf.ps.gz}  }}

\bigskip
{\cl {\bf {Marc Davis}}}

{\cl {Departments of Astronomy and Physics}}
{\cl {University of California, Berkeley, CA. 94720, U.S.A}}
{\cl {marc@coma.berkeley.edu}}

\medskip

{\cl {\bf {Adi Nusser}}}

{\cl {Institute of Astronomy, Madingley Rd, Cambridge, CB3 0HA, England}}
{\cl {adi@ast.cam.ac.uk}}

{\cl {and}}

{\cl {\bf Jeffrey A.\ Willick}}

{\cl {Department of Physics, Stanford University, Stanford, CA 94305-4060}}
{\cl {jeffw@perseus.stanford.edu}}

\bigskip

{\cl {\bf ABSTRACT}}
We consider a measure of the peculiar
velocity field derived from the Mark III compilation of 2900 spiral
galaxies (Willick \etal 1996b), using an analysis method that is substantially
free of bias (Nusser \& Davis 1995).  We expand the velocity
field in a set of  orthogonal, smooth modes, reducing the data to a set
of 56 coefficients fitted to a maximum redshift of $6000 \kms,$
and maximum spherical harmonic of $l=3.$ The radial resolution of the
modes degrades 
with redshift, from $800 \kms$ locally to $3000 \kms$
at $4000 \kms$ redshift.
Equivalent mode coefficients can be computed for the gravity field derived
from any whole-sky redshift catalog of galaxies, such as the \iras 1.2
Jy survey (Fisher \etal 1995).  Given the coefficients of the
expansions, one can compare the velocity and gravity fields on a galaxy
by galaxy basis, or on a mode by mode basis.  Detailed
comparison shows the two independent fields to be remarkably aligned in
general.  There are, however, systematic discrepancies in the fields
that lead to considerable coherence in the residuals between them.  These 
residuals take the form of a dipole field in the
LG frame that grows with distance; it is 
not consistent with a bulk flow residual.  

 We perform a likelihood analysis in the mode-mode comparison to
determine which value of $\beta\equiv \Omega^{0.6}/b$  
for the \iras gravity field is the best fit to the Mark
III velocity field, considering the errors and covariance in both
the velocity and gravity coefficients.  
We find that the most likely value 
lies in the range $\beta=0.4$--$0.6.$ However, in contrast
with results we obtain using simulated galaxy catalogs,
the $\chi^2$ per degree of freedom for the 
fit is well in excess of unity, primarily because of the
coherent dipole residuals at $cz\simgt 3000 \kms.$ Thus,
despite the general alignment of the
Mark III velocity and \iras gravity fields, they do
not agree in detail, precluding a firm determination
of $\beta$ from these data sets at present.  
The method is
capable of measuring $\beta$ to an accuracy of 10\%, but 
without understanding these systematic discrepancies, we cannot 
infer a value of $\beta$ from these data.

\bigskip\noindent
{\it Subject headings:}\ cosmology ---  dark matter ---
galaxies: clustering --- galaxies: formation --- gravity --- large-scale
structure of the universe

\bigskip
 
\vfill\eject

{\cl {\bf 1. INTRODUCTION}}

\medskip

The observed large scale flows of galaxies are a direct probe
of the nature of the underlying dark matter in the Universe.
Under the assumption that the large scale structure has formed via
gravitational amplification of small initial density fluctuations,
the observed velocity field 
yields a prediction for the
present matter-density fluctuations given an assumed value for $\Omega$
(Nusser \etal 1991). 
Furthermore,  quasi-linear gravitational instability theory can
be used to reconstruct the primordial density fluctuations from
the observed flows. This enables us to address fundamental questions
about the nature of the initial fluctuations and their relation
to $\Omega$. For example, Nusser \& Dekel (1993) have demonstrated that
unless $\Omega$ is around unity the initial density fluctuations were
non-gaussian. 

The largest virialized systems in the Universe are the rare, rich
clusters of galaxies, with typical sizes of less than 1
$h^{-1}$ Mpc.  The large scale flows, on the other hand, are motions
with coherence length of order 10 - 40 $h^{-1}$ Mpc, more than an
order of magnitude larger than the rich cluster sizes.  Measurements
of mass by means of the alignment which is expected between velocity
and gravity fields on these large scales is thus quite distinct from
the virial, or hydrostatic, or weak lensing analyses that can be
applied to the clusters.  A trend of increasing mass to light ratio is
well documented to exist with increasing scale size of the measurement
(Ostriker \etal 1974, Davis \etal 1980).  Rich clusters fall on this
trend, with an inferred density parameter of order $\Omega_0 \approx
0.1-0.2$, consistent with the baryon fraction in the clusters (e.g.
White \etal 1993).  A fundamental question is whether the trend of
$\Omega_0$ versus scale asymptotes at this value, or continues to
climb upward on larger scales.  If galaxy formation is more efficient
in denser regions or if galaxies have some degree of velocity bias, or
if the dark matter is too hot to fully fall into the cluster potential
wells, then we might expect to measure a larger value of $\Omega_0$ on
the scale of the large-scale flows.  Considerations of an inflationary
early Universe have traditionally led to the expectation of
$\Omega_0=1$ (although see e.g.  Liddle and Lyth 1993, Bucher and
Turok 1995); our confidence in such speculations would be considerably
enhanced if someone could actually produce compelling evidence, on any
scale, to support such a high density.

On large scales, virial equilibrium does not apply, but one can still
derive estimates of the mass density, now based on flow models, either
linear or nonlinear.  The earliest detailed flow models were based on
Virgocentric infall (e.g.  Aaronson \etal 1982, Tonry and Davis 1981).
These models assumed a smooth background and spherical symmetry for
the local supercluster, and this restrictive symmetry allowed the
incorporation of nonlinearity in the relationship between velocity and
density.  We now know that spherical symmetry is a rather poor
approximation for the infall to any cluster, but linear theory
analysis, which assumes no special symmetry, remains a valid tool.
The large scale flows are almost certainly the result of the process
of gravitational instability (e.g. Dekel \etal 1993, 
but see Babul \etal 1994), with the
overdense regions attracting material, and 
underdense 
regions repelling material.  Initial conditions in the early universe
might have been somewhat chaotic, so that the original peculiar
velocity field (i.e.  deviations from Hubble flow) was uncorrelated
with the mass distribution, or even contained vorticity.  But those
components of the velocity field which are not coherent with the
density fluctuations will adiabatically decay as the Universe expands,
and so at late times one expects the velocity field to be aligned with
the gravity field, at least in the limit of small amplitude
fluctuations.  In this paper we shall attempt to make this comparison
directly, using the best presently available data for both the
velocity and gravity fields.

The subject of large scale flows has been extremely active of late, and
two excellent reviews are now available (Dekel 1994; Strauss and Willick 1995).
The observational situation is rapidly improving; in this paper we shall
consider the velocity field derived from a sample of 2900 spiral galaxies
with Tully-Fisher measurements (Willick \etal 1995; Willick \etal 1996a,b).
The basic 
principle is to use late-time linear theory to relate
the measured velocity field to the  gravity field as predicted
from the observed galaxy distribution.
Given complete knowledge of the mass fluctuation field $\delta_\rho({\bf r})$
over all space, one could trivially write the
fluctuations in the gravity
field ${\bf g(r) } $ as  
$$
 {\bf g}({\bf r}) =  G\bar{\rho} \int d^3 {\bf r'} \delta_\rho({\bf r'})
{\bf  r' - r \over |r'-r|^3 } \ , \eqno(1)
$$
where $\bar\rho$ is the mean mass density of the Universe.  
The great simplification of late time linear theory is the intuitive relation
between the gravity field {\bf g} and the velocity field ${\bf v_L}$,
namely   ${\bf v_L} = {\bf g}~ t$ where 
the only possible time $t$ is the Hubble time.  The correct expression
(Peebles 1980) is 
$$ {\bf v_L(r)} = {2 \over 3 H_0 \Omega^{0.4}} {\bf g(r)} \ . \eqno(2)  
$$ 
If the galaxy distribution at least
approximately traces the mass on large scale, with linear bias $b$
between the galaxy fluctuations $\delta_G$ and the mass fluctuations
(i.e. $\delta_G = b \delta_\rho$),
then from (1) and (2) we have  
$$ 
{\bf v_L(r)} = {H_0 \beta \over 4 \pi {\bar n}}
 \sum_{i} {1\over\phi(r_i)} {\bf  r_{\rm i} -r  \over | r_{\rm i} -r|^3 }
+ {{H_0 \beta }\over 3}{\bf r}\ , \eqno(3) 
$$ 
where $\bar n$ is the true mean galaxy density in the sample, $\beta \equiv
\Omega^{0.6}/b$, and where we have replaced the integral over space
with a sum over the galaxies in a catalog, with radial selection
function $\phi(r) $ .  ($\phi(r)$ is
defined as the fraction of the luminosity distribution function observable
at distance $r$ for a given flux limit; see e.g. (Yahil \etal 1991).) 
 The second term is the correction
for the uniform component of the galaxy distribution which would exactly
cancel the first term in the absence of clustering within the survey volume.  
One problem with Equation (3) is that the sum is to
be computed in real space, whereas the galaxy catalog exists in redshift
space. Furthermore, the integral is to be computed over all space, but the sum
over the catalog must be truncated at some finite radius.
Note that the result is insensitive to the value of $H_0$, as
the right hand side has units of velocity.  We shall henceforth quote
all distances in units of $\kms$.  The density $\bar n$ and selection
function $\phi(r)$ can be estimated from the redshift catalog used in
the analysis, so that comparison of the measured velocity field to the
predicted velocity field ${\bf v_L(r)}$ gives us a measure of $\beta$.
The degeneracy between $\Omega$ and $b$ can be removed in principle by
nonlinear effects (Nusser \etal 1991, Dekel \etal 1993), but in practice 
one should not be optimistic because, among other problems, nonlinear 
bias will enter at the same order as nonlinear velocity effects. 

The direct comparison of the peculiar velocity of selected galaxies
versus the gravity predicted from Equation (3) is fraught with
difficulty. Distances to individual galaxies are typically uncertain at
the level of 15-20\% and are furthermore subject to considerable
Malmquist bias.  To compute the gravity field, one needs a nearly whole
sky redshift catalog of galaxies.  To date the best available catalog
is the 1.2 Jy flux limited \iras catalog of 5300 galaxies (Fisher \etal
1995), which we shall use in this analysis.  The \iras predicted
gravity field has substantial uncertainty because of the dilute
sampling of the density field. Furthermore, the predicted velocities
 are subject
to large covariance, because the \iras predicted field is quite
non-local.  Flows within the turnaround radius of large  clusters are
multivalued, so that in projection a given redshift can correspond to
three separate distances. The examination of numerical
simulations demonstrates that coherent shell crossing is a common
occurrence near dense centers: the redshifts of a set of
points can be larger (smaller) than the redshifts of nearby points with 
smaller (larger) distances. In this case, essential information
has been lost and recovery of the correct distances from redshift
information alone is impossible.

Several comparisons of velocity 
and gravity fields have been made with older
datasets (e.g. Hudson 1994; Kaiser \etal 1991; Yahil 1988; Strauss and Davis
1988) but these analyses were all meant to be preliminary and none of 
their conclusions were  compelling. To date nobody has
included a proper treatment of the correlated noise in the analysis.  
Shaya \etal (1995) have 
compared the peculiar velocities of a set of 298 spiral galaxies within
a redshift of 3000 $\kms$ to the gravity field derived by least action
analysis of the 1138 mass tracers dervived from the 
{\it Nearby Galaxies Catalog} (Tully 1988).  
Their preliminary analysis gives a value $ \Omega_0 = 0.2 \pm 0.2 $, 
but they have not included covariance in the
errors of the predicted mass model, nor have they realistically included
possible influence from the mass distribution at redshifts $z>3000 \kms,$ 
 nor is it clear how well their analysis
compares with the linear theory techniques used by others.

Because of the covariance of the \iras gravity field 
a comparison of the mass fluctuations as inferred from the flows versus
the \iras density field has considerable merits.  
Assuming the flow is irrotational, the POTENT algorithm (Dekel \etal 1990) 
reconstructs a three dimensional flow from the observed
radial flow field, and then computes derivatives of this flow to
infer the underlying mass field (Nusser \etal 1991).  This then
can be compared directly to the \iras density field (or to any
other survey).   Such a comparison was done by 
Dekel \etal (1993),
using an earlier version of the \iras catalog, and the Mark II
catalog of peculiar velocities for 493 objects (Faber and
Burstein 1988).  On the basis of this density-density comparison,
Dekel \etal (1990) derived $\beta = 1.3 \pm 0.3$.  
More recently, Hudson \etal\ (1995) have compared 
POTENT mass densities reconstructed from the Mark III data 
(Willick \etal 1995, 1996a, 1996b) 
with the optical density field of Hudson (1994),
finding $\beta=0.74\pm 0.13.$
A new analysis
using the 1.2 Jy \iras survey and the Mark III peculiar velocity
catalog is in preparation (Dekel \etal 1996).

Given the significance of the questions raised, all viable alternative
algorithms should be pursued to extract the implications of the large
scale flows.  In this paper we shall compare the observed radial
velocity field with the radial gravity field inferred from the 1.2 Jy
IRAS redshift survey (Fisher \etal 1995).  We shall use the method of
orthogonal mode expansion as described by Nusser and Davis (1995),
which we shall henceforth label  as the Inverse Tully-Fisher (ITF) method.  
A disadvantage of the density-density comparison is that the
mass-fluctuations which are derived by POTENT increase with decreasing
$\beta$. According to the quasi-linear approximation (Nusser \etal 1991)
employed by POTENT, the mass-fluctuation field inferred from the
velocities diverges for low $\beta,$ thus invalidating the density-density
comparison. Therefore the density-density comparison
can not formally rule out very low values of $\beta$. However, 
the fluctuations in the velocity field 
predicted from the observed distribution
of galaxies are smaller for lower $\beta$ and the applicability of 
the linear approximation is guaranteed. Therefore, the velocity-velocity
comparison provides a better tool for assessing the agreement
between the observations for low $\beta$.

This paper closely follows the methods described in Nusser and Davis
(1994, hereafter ND94) and in Nusser and Davis (1995) (hereafter
ND95), with some additional refinements developed since then (Davis
and Nusser 1995).  In Section 2, we review the essentials of
these methods and describe our choice of basis functions.  We also
discuss the procedure for fitting the modes to both the velocity and
gravity fields.  As has been the case in the past, mock catalogs
constructed from $N$-body simulations are essential for debugging and
calibrating the methods.  This is especially so for our application,
since the entire analysis is performed in essentially pure redshift
space.  In Section 3, we show a few new details on
the tests applied to the mock catalogs.  Section 4 gives details of
the comparison of the Mark III velocity field with the \iras inferred
gravity field, on a point by point basis, as well as on a mode by mode
basis.  A detailed likelihood analysis is presented in Section 5,
where the needed covariance matrices are derived as well.
Section 6 presents a discussion and final conclusions are given in
Section 7.

\bigskip

{\cl {\bf 2. Reconstruction of Peculiar Velocities }}

In this section we outline the methods of ND94 and ND95 for
deriving the smooth peculiar velocities of galaxies from
an observed distribution of galaxies in redshift space and, independently,
from a sample of spiral galaxies with measured circular velocities 
$\eta\equiv {\rm log} \Delta v -2.5$ and apparent magnitudes $m$.  
Any of the available methods (e.g. Yahil \etal 1991, Fisher \etal 1995b)
for generating \iras predicted peculiar velocities can be used. 
However, the method of ND94
is particularly convenient, as it is easy to implement, fast,
and requires no iterations. Most importantly, this redshift space
analysis closely parallels the ITF estimate described below. 
Since we shall use this analysis in what
follows, we next present  a very brief summary of the
method.

\medskip
{\cl {\it 2.1 
Peculiar Velocities from the Distribution of Galaxies in Redshift Space}}

Let  $\vv(\vs)$ be the comoving peculiar velocity
relative to the motion of the Local Group (LG) and $\vs$ the comoving redshift 
space coordinate. (We use $s= cz/H_0$, noting that higher order
terms in $z$ have a  negligible influence on the derived velocity
field).
To first order, the peculiar velocity is irrotational
in redshift space and it can be derived from a potential function, 
$\Phi(\vs)$, defined by 
$\vv(\vs)=-\vnabla\Phi(\vs)$. 
Let $\delta_0(\vs)$ be the fractional fluctuation
of the density field traced by the distribution of galaxies in 
redshift space.
If we expand the angular dependence of $\Phi$ and $\delta_0$ in terms
of spherical harmonics (Fisher, Scharf \& Lahav 1994, Fisher \etal 1995b) 
in the form, $\Phi(\vs)=
\sum_{l=0}^{\infty}\sum_{m=-l}^l \Phi_{lm}(s)Y_{lm}(\theta,\varphi)$
and similarly for $\delta_0$, then, to first order,
$\Phi_{lm}$ and $\delta_{0lm}$ satisfy,
$$
{1\over {s^2}} {\dd \over {\dd s}}
\left(s^2 {{\dd \Phi_{lm}} \over {\dd s}}\right)
-{1 \over {1+\beta}}{{l(l+1) \Phi_{lm}} \over {s^2}}
={\beta \over {1+\beta}} \left(\delta_{0lm} - {1\over s}
{ {\dd \ln\phi} \over {\dd \ln s} } { {\dd \Phi_{lm}} \over {\dd s}}\right).
\eqno (4)
$$
where $\phi$ is the selection function of the sample.  As emphasized by
ND94, the solutions  to (4) for the monopole ($l=0$) and the dipole
($l=1$) components of the radial peculiar velocity in the LG frame 
are uniquely determined by specifying vanishing velocity at the origin.  
That is, the radial velocity field at redshift $\bf s$,  when expanded to
harmonic $l \le 1$, is not influenced by material at redshifts greater
than $\bf s$.

As an estimate of the field $\delta_0(\vs)$ from the discrete distribution 
of galaxies we consider,
$$
\delta_0(\vs)= 
{1\over {(2\pi)^{3/2}\nb \sigma^3(s)}}\sum_i {{1}\over {\phi(s_i)}} 
\exp\left[-{{\left(\vs - \vs_i\right)^2}\over
{2\sigma^2(s)}}\right] -1 \quad . \eqno (5)
$$
We emphasize that the coordinates ${\bf s}$ are in {\it
observed redshift} space, expanded in a galactic reference frame.  The only
corrections from pure redshift space coordinates is the
collapse of the fingers
of god of the known rich clusters prior to the redshift space smoothing
(Yahil \etal 1991).
The density field is here smoothed by a gaussian window 
with a redshift dependent width, 
$\sigma(s)$, which is
proportional to the mean particle separation, i.e, $\sigma(s) \propto
(\nb\phi(s))^{-1/3} $, where $\nb$ is the mean density of the \iras sample.
The purpose of the smoothing is to limit the discreteness 
noise of the gravity field
constructed from the dilute redshift sample.
This particular smoothing ensures that the error in the field 
(5) due to Poisson noise is constant in
space for the case of a uniform distribution of galaxies.  A more
elaborate realization  of the density field based on Wiener filtering
is described by Fisher \etal (1995b), with results very close to those
obtained here.  Weighting  the galaxies in (5) by
the selection function evaluated at their redshifts rather than the
distances which are unknown yields a biased estimate for
the density field and gives rise to Kaiser's rocket effect (Kaiser 1987).  
However, this bias should be compensated by the term involving the selection
function in (4).

\bigskip
{\cl {\it 2.2  Peculiar Velocities from the Inverse Tully-Fisher relation:}}
{\cl {\it Expansion of the Fields in Orthogonal Functions}}

Given a sample of galaxies with measured circular velocity
parameters, $\eta_i$, apparent magnitudes, $m_i$, and redshifts,
$z_i$, the goal is to derive an estimate for the smooth
underlying peculiar velocity field.  We assume that the circular
velocity parameter, $\eta$, of a galaxy is, up to a random
scatter, related to its absolute magnitude, $M$, by means of a
linear inverse Tully-Fisher (ITF) relation, i.e.,
$$
\eta=\gamma M + \eta_0 .  \eqno (6)
$$
 
The advantages of inverse TF methods were first described in 
detail by Schechter (1980) and Aaronson \etal (1982). 
Most critically, samples selected by magnitude, as most are, will
not be plagued by  Malmquist bias effects when analyzed in the inverse
direction.
Methods based on an inverse relation generally require a velocity model 
(Schechter 1980, Aaronson \etal 1982, Lynden-Bell 1991, Lahav 1991) which 
can be parameterized in terms 
of observable redshifts. ND95 consider a very general model and 
write the absolute magnitude of a galaxy, $M_i = M_{0i} + P_i$, where
$M_{0i} = m_i + 5{\rm log}(z_i)-15$ and $P_i = 5{\rm log}(1- u_i/z_i)$, 
where $m_i$ is the
apparent magnitude of the galaxy, $z_i$ is its redshift 
in units of $\kms,$
and $u_i$ its radial peculiar velocity in the LG frame.  
In general, one can write the
function $P_i$ in terms of an expansion over orthogonal functions,
$$
P_i =\sum_{j=0}^{j_{max}} \aj \Ft^j_i \eqno (7)
$$
with orthonormality conditions,
$$
\sum_{i=1}^{N_g} \Ft^j_i  \Ft^{\jp}_i= \delta_K^{j , {\jp}} , \eqno (8) 
$$
with the zeroth mode defined by $\Ft^0_i=1/\sqrt N_g$, where $N_g$ is 
the number of galaxies in the sample. The zeroth mode
describes a Hubble-like flow in the space of the data set which is clearly
degenerate with the zero point of the ITF relation.
Here we arbitrarily set ${\tilde \alpha}^0=0$. 
The best fit parameters, $\aj$, the slope, $\gamma$, and  the zero point
$\eta_0$, are found by  minimizing the $\chi^2$ statistic
$$
\chi^2 =\sum_i
  {{\left(\gamma M_{0i}+\gamma P_i+\eta_0-\eta_i\right)^2}\over 
{\sigma_{\eta}^2}}
. \eqno (9)
$$
where $\sigma_{\eta}$ is the $rms$ scatter in $\eta$ 
about the ITF relation.
With our choice for the zeroth mode, the solution  for $\aj$ is,
\def\smoo{\sum_i M_{0i}^2}
\def\smeoo{\sum_i M_{0i}\eta_i}

\def\sumup{\sum_{i,{\ip}}M_{0i}\eta_{\ip}\left(1+N_g\sum_j \tj^j_i 
\tj^j_{\ip}\right)}
\def\sumdown{\sum_{i,{\ip}}M_{0i}M_{0\ip}\left(1+N_g\sum_j\tj^j_i 
\tj^j_{\ip}\right)}
\def\aq{{\sumdown -N_g\smoo}}
\def\bq{{\sumup-N_g\smeoo}}
$$
\eqalign{\aj&= 
-{1\over \gamma}\sum_i \left(\gamma M_{0i}-\eta_i\right)\tj^j_i , \cr
\gamma&={{\bq}\over {\aq}} .  \cr} \eqno (10)
$$

	As described in ND95 the choice of radial basis functions for
the expansion of the modes can be made with considerable latitude.  The 
functions should obviously be linearly independent, and close to orthogonal
when integrated over volume. They should be smooth and close to a complete
set of functions up to a given resolution limit.   ND95  chose spherical
harmonics $Y_l^m$ for the angular wavefunctions and the 
derivatives of spherical Bessel functions for the radial basis functions, 
motivated by the desire to use functions which automatically satisfy
potential theory boundary conditions at the origin and the outer boundary.
That is, they chose 
$$
P(z,\theta,\phi)=\sum_{n=0}^{n_{max}}\sum_{l=0}^{l_{max}}\sum_{m=-l}^{m=l}
{a_{nlm} \over z}\left(j^{\prime}_{l}\left(k_nz\right)-c_{l1}\right)
Y_{lm}\left(\theta,\phi\right) \quad .
\eqno (11)  $$
The constant $c_{l1}$ is non-zero for the dipole term only and ensures that
$P=0$ at the origin, and is non-zero at the outer boundary.
Details of how the orthogonalized functions $\tilde F_i^j$ are derived
from this expansion are given in ND95.

To keep the expansion of the basis functions compact, it is
sensible to match the resolution of the wavefunctions to the
spatial distribution of the Tully-Fisher data.  Since the
precision of the inferred peculiar velocities is a  constant
fraction of the distance to any galaxy, spherical Bessel
functions, which oscillate uniformly with their argument, are not
the optimal choice of wavefunction.  
Furthermore, the mean interparticle spacing of galaxies in flux
limited redshift catalogs, from which we infer
the density and gravity fields, 
can increase even more rapidly than linearly
with distance. For example, for  $z\gg 6000 \kms,$  
the mean interparticle spacing within the 1.2 Jy \iras survey
increases as $z^{1.6}$.
Therefore, to
make optimal use of the available information from the ITF field, 
we expand the modes of $P_i$ in terms of a transformed radial coordinate
that 
decreases the oscillation periods of the Bessel functions
roughly according to the distribution of data as a function of redshift.
A convenient, but by no means unique choice is 
$$ 
y =  \left({\rm ln}\left[1 + (z/z_*)^2\right]\right)^{1/2}  ~. \eqno (12)
$$
For an increment $\Delta y$, 
this function has resolution $\Delta z = (dy/dz)^{-1}\Delta y$ that is
constant on small
scale and $\Delta z \propto z \ln(z)$ on large scale.  We set
$z_* = 1000 \kms.$
The radial wavenumbers for each $l$ value have been chosen for the lowest
values of $k_n$ such that
$$j_{l-1}(k_n y_{max}) = 0 ~. \eqno (13) $$
This choice does not restrict the value of 
the expanded velocity field near the boundary $y_{max}$.
As we shall see, transforming to the variable $y$ results in a significant
reduction in the value of the $\chi^2$ of the fit 
relative to the expansion in terms $z$.
Given these choices of radial variable and values of $k_n$, the resulting
wavefunctions for $l=0,4$ are plotted in Figure 1. Note how the wavelength
of the modes  increases with distance. 
All of the modes are forced to
be orthogonal to the $P=constant$ mode 
when summed over the set of points in 
the Mark III catalog, and this is accomplished by  small additive constants. 
However,  the $n=1,$ $l=0$ mode is 
positive definite over the volume, and to make
it orthogonal to the $P= constant$ mode would require a substantial additive
offset, which would then violate the proper boundary condition at the
origin.  We therefore delete this mode from inclusion in our expansion set.

	As a demonstration of the use of the variable $y$, 
we tested the Mark III catalog which lists 2237 galaxies
within $s= 6000 \kms$, details of which will be described below.
We took the linewidth scatter to be $\sigma_\eta = 0.05$
(Willick \etal 1995).
For no velocity model, (i.e. setting all coefficients $\alpha^j =0$), 
the scatter diagram
of the Tully-Fisher regression of observed versus predicted linewidths
 $\eta$ yields  $\chi^2= 3516$.  Fitting a flow
model of 69 degrees of freedom ($l_{max}=3$, $n_{max}=4$) in terms of
radial wavefunctions $j'_l(k_n z)/z$ leads to a much reduced $\chi^2 =
2667$, while doing the same fit of 69 modes in terms of $j'_l(k_n
y)/z$ gives $\chi^2 = 2644$, a modest but significant
improvement. Further tests on simulations with smaller
$\sigma_\eta$ demonstrate that 
the effect of the coordinate transformation 
is considerably more pronounced.
There is clearly room for refinement here, but this choice of radial
variable is sufficient for our purposes, especially since we shall use
identical basis functions for the expansion of the velocity and gravity
fields.

\bigskip

\noindent{\it 2.2.1 Choosing the resolution limit}

	In the comparison of the Mark III velocity and \iras  gravity fields
discussed below, we shall use the Mark III sample only within  a redshift limit
of $6000 \kms.$  This corresponds to setting the outer boundary condition
at $y_{max} = 1.9$.  
If we consider that the number of peaks in the $n$th radial
wavefunction is $n$, then the 
effective resolution $R$ at any redshift is 
$R(z) \approx \left({dy \over dz}\right)^{-1} {y_{max} \over n}$.
For $n_{max} = 4$, $z_{max} = 6000 \kms,$ and
$z_* = 1000 \kms,$ the resulting limiting radial resolution is $680 \kms$ at
$z = 1000 \kms,$ and $3400 \kms$ at $z = 4000 \kms.$
 
The angular resolution for a spherical harmonic function $Y_{lm}$
is $180/l$ degrees. For redshift $z$, this corresponds to a
maximum spatial resolution of $R \approx {\pi z / l_{max} }$.
If we desire to have the angular and physical resolutions
approximately equal at $3000 \kms,$ where the radial resolution is
approximately $2400 \kms,$ then we should select $l_{max} = 3.9
\approx 4$.  We shall in fact limit $l_{max} $ to 3, in order to
reduce the number of degrees of freedom.  The result is that our
radial resolution is slightly higher than our transverse
resolution, but by our choice of radial variable $y$, the ratio
of transverse to radial resolution is approximately constant with
distance, which is a desirable feature. 

The choice of the variable $y$ is governed by the desire to have a mode
resolution that is a fixed fraction of the resolution of the TF data, but as 
long as the resolution is below that of the intrinsic data, we can be assured
that both the smoothed velocity and gravity fields have identical 
resolution.   We have chosen a mode resolution corresponding to 
the spatial frequency of the fourth dipole mode: $k = {4\pi\over y_{max}}$
That is, we include the first 4 radial modes that satisfy the outer
boundary condition, equation (13), for the harmonic $l=1$.
We  use this limiting cut in wavenumber as a limit for the other spherical 
harmonics.  Thus to the same limiting value of $k,$ we should
include 4 modes for $l=0$, the first of which is discarded, 3 modes for $l=2$,
3 modes for $l=3$, 2 modes for $l=4$, 2 modes for $l=5$, and 1 mode for $l=6$.
However, tests based on the efficiency of the
decrement in the $\chi^2$ value of the fit demonstrated that
a cutoff in $l$ is desirable here. Thus on top of the
sharp cutoff in $k$-space we eliminate modes with harmonic index
exceeding $l_{max} 
= 3,$ 
as mentioned above.
This yields $3\times 1 + 4\times 3 + 3\times 5 + 3\times 7 +5
= 56$ modes in total.
The Mark III velocities within a limiting redshift of $6000 \kms$ 
were then derived from the ITF method. 
The peculiar velocities as predicted from the \iras real or mock
catalogs were expanded in the same ITF 
basis functions, as described next.

\bigskip
\noindent{\it 2.2.2  Expanding the \iras  velocity field}


Using  the machinery for computing a gravity field described in \S 2.1, 
one can generate a linear theory predicted peculiar velocity
$v_L$ for any point in space as a function of its redshift for
any value of $\beta$.  We must ensure that the smoothing scales
of the ITF  and \iras predicted peculiar velocities are matched
to the same resolution. Therefore we expand $v_L$ in terms of the
modes used in the velocity model (12).  Because of the
orthonormality, we can write the mode coefficients as
$$
\alpha_{iras}^j = 5 \sum_i {\rm log}\left(1 - {(v_{Li} - H' z_i)\over z_i} 
\right) \Ft^j_i
~ , \eqno (14)
$$
where the $H'$ term is a correction for Hubble flow and the summation
index $i$
is restricted to be
over the positions of the same galaxies in the ITF expansion.

This procedure will filter out higher frequency modes that are not
described by the resolution of our basis functions.  However, it
is important to recall that the modes are chosen to be orthogonal
to the mode $P_i = $ constant, which would describe a smooth
Hubble flow.  In the fitting for the ITF modes, pure Hubble flow
is absorbed into a shift of the zero point $\eta_0$ and the
orthogonality is ensured.  Within a given set of test points
occupying a volume smaller than that used to define the gravity
field, it is possible for $v_L$ to have a non-zero value of
Hubble flow $H'$, which must be removed from $v_L$ before we
tabulate the mode coefficients.  That is,  we tabulate the mean
Hubble ratio
$$
H' = {1\over N} \sum_i {v_{Li} \over z_i}    \eqno (15)
$$
and subtract it from the predicted field $v_L$.  This ``breathing
mode" which mimics a Hubble flow is not
trivial in amplitude, and can be a 10\% correction on the effective Hubble
constant within simulated catalogs.

\bigskip
{\cl {\bf 3. Tests with Mock Catalogs}}

ND94 and ND95 tested the velocity reconstruction methods and
their biases using mock catalogs drawn from an $N$-body simulation
of a  CDM, ($h=0.5$, $\Omega_0=1$, and  $\sigma_8=0.7$) universe. The
simulation had the disadvantage of having low amplitude large
scale flows and therefore was not fully suitable for our
purposes.  Furthermore, the algorithm for attenuating the rather
large small-scale incoherent  velocities seen in the simulation
resulted in too much reduction of their amplitude.  Here we
provide more powerful tests of the methods using mock Mark III
and 1.2Jy \iras catalogs which are extracted from a pure $n=-1$, 
$\Omega_0 = 1$ simulation of $10^6$ points kindly supplied by
S. White.  We use a normalization $\sigma_8 = 0.7$, which then
determines that the size of the box is $L = 235h^{-1}$ Mpc.  Such
a  simulation has the desirable feature of having 
very high-amplitude large 
scale flows.  In this simulation, as in the CDM simulation used
in ND, the one-dimensional  small scale (few Mpc scale)
velocities are far in excess of that observed for the real
universe ($\approx 140-200 \kms;$ see Miller \etal 1996) 
and have
to be smoothed somehow.  Instead of replacing the $N$-body peculiar
velocities with Gaussian smoothed velocities as we have
done in previous work (Fisher \etal 1993),
we replace the original $N$-body velocities with the average of a
Gaussian smoothed field (with smoothing length $\sigma = 100 \kms$) and the
original field.  Thus we reduce the one dimensionsal RMS
amplitude of the very small scale
velocity field $\sigma_u$ from $400$ to $200 \kms$ while  keeping the
large scale velocities unchanged.  This is of course not a
self-consistent procedure, but better approximates the observations
and the expected  effects of
possible velocity bias, or of baryon dissipation within extended
dark matter halos.  In dense regions, with $\bar\delta > 600$,
we collapse the redshift space fingers of god,
as is done in the centers of real clusters for both the Mark III
and \iras catalogs.

In order to generate realistic mock catalogs from the simulation,
a particle moving at $600 \kms,$ with small local shear, and
which is not in a dense region is chosen as the central observer.
A mock full sky 1.2Jy \iras catalog is generated from the
simulation by assigning luminosities to particles according to
the luminosity function of the 1.2Jy \iras survey. This procedure
results in 5021 ``galaxies" out to a redshift of $180{\rm
h^{-1}Mpc}$ from the ``observer''.

To select mock Mark III catalogs, we follow the procedure
described in ND95, with the following modifications: we extract a
nonuniform subset of points from the mock \iras catalogs, with
galaxies at $|b| < 18^\circ$ heavily selected against, and with
objects in the hemisphere of the microwave dipole anisotropy
vector having twice the probability of selection as objects in
the anticenter direction.  Each point within the mock catalog has
an absolute magnitude drawn from the luminosity function of the
1.2 Jy \iras\  catalog.  The selected points in the mock Mark III
sample are given a linewidth $\eta$ drawn from a  linear
Tully-Fisher  law with $\eta_0 = -2$ and $\gamma=-0.1$, and with
gaussian scatter $\sigma_\eta = 0.05$, a value we find to
closely fit the observed scatter in the real Mark III catalog.
The mock Mark III data
thus consist of a position on the sky, a redshift, apparent
magnitude, and linewidth for each galaxy; the true peculiar
velocity of each ``galaxy" is used only to test the reliability
of the reconstruction.  A much more elaborate construction of
mock Mark III catalogs is described by Kolatt \etal (1996).  We
have tested our procedures on this catalog for which we have no
more difficulty recovering the true underlying field than with
the mock catalogs described here.

With these two catalogs, we can generate mock ITF velocity fields
as described above, and mock \iras velocity fields using the technique
of ND94 described above.  By simply changing the random number
seed, we have generated different realizations of the mock Mark
III sample for the same points, and different mock \iras catalogs
from the same central observer within the $N$-body simulation.

To generate the mock \iras velocity field by means of the method
described in Section 2.1, we first use the unbiased
``galaxy'' distribution out to $18000 \kms$ to generate a smooth
density field.  We fix the Gaussian smoothing length $\sigma(s)$ to
be the mean particle separation at the redshift $s$.  For the 1.2 Jy
\iras catalog, this smoothing is approximately $\sigma(s)=345 \kms$
at $z=1000 \kms,$ $\sigma(s)=570 \kms$ at $z=3000 \kms,$ and
$\sigma(s)=980 \kms$ at $z=6000 \kms,$ considerably smaller than
the mode filtering we shall apply to the ITF velocity and \iras
gravity fields, but large enough to filter out many of the nonlinear
structures in the mock \iras catalog.  The density field and
resulting velocity field (equations 4 and 5) are then expanded in
spherical harmonics up to $l_{max}=12$, although in fact a lower
cutoff at $l=4$ is sufficient.  Using this relatively high resolution
velocity field based on the mock \iras catalog, we generate a linear
theory prediction for each point in the mock Mark III catalog.

We use the 56 mode expansion to minimize the scatter of the
Tully-Fisher relation in the mock Mark III catalog; for the 1733
points in the mock catalog within $s=6000$ km/s,
the $\chi^2$ of the fit is reduced from 2264 with no flow, to 1900 with
a 56 mode flow model. 
Note that the reduced $\chi^2$ is somewhat greater than unity in
spite of our precise knowledge of the scatter in $\eta$.
The slightly high $\chi^2$ is simply due to components of the
velocity field with higher spatial frequency than the resolution of
our mode expansion; however, increasing the number of modes in the expansion
reduces $\chi^2$ relatively slowly.  If the $\eta$ scatter were 
smaller, more degrees of freedom  would be worthwhile. 
 We recover best values of $\eta_0 = -1.94$ and
$\gamma = -0.0965$, only slightly off their input values of $-2.0$, $-0.1$.

We find it most useful to present the results of the flow model
pictorially. In Figure 2 we present three redshift slices of the
ITF velocity field $u_{itf}$ derived for the mock catalog using
the 56 mode expansion.  The plots are in galactic coordinates
where stars (circles) represent mock galaxies outflowing
(inflowing) in the LG frame, with sizes proportional to the flow
amplitude.  In Figure 3 we show $u_{itf}$ projected onto the
Supergalactic Plane.  Note that although the data are irregularly
distributed, the velocity field is very smooth, by construction.
These plots, like all the plots in this paper, are in the Local
Group frame.  Note the strong reflex dipole signature which
dominates the large scale flow.

The real advantage of working with mock catalogs derived from
self-consistent $N$-body simulations is that for each point we
know the true peculiar velocity $u_{true}$, including all the
nonlinear, small scale effects.  Figure 4a shows the true,
unfiltered velocity field $u_{true}$ of the mock Mark III points,
projected to the supergalactic plane, while Figure 4b shows the
true field after filtering by the ITF mode functions
$u^f_{true}$, Equation (14).  Note how the region of substantial
shell crossing shown near the clusters of the upper left section
of Figure 4a is largely eliminated in the filtered version of the
field, Figure 4b.  Figure 4c shows the \iras derived, unfiltered
velocities $u_{iras}$, while Figure 4d shows the \iras derived
velocities after filtering, $u^f_{iras}$.  The $u_{true}$ field
has a dipole signature on large scale, and considerable complex
structure around the dense knots; the $u_{iras}$ field is
considerably smoother and cleanly separates the infalling and
outflowing zones, particularly in the upper left portion of
Figure 4c, as expected from a Zeldovich approximation
methodology.  Note that Figures 3, 4b, and 4d, the smoothed
versions of the ITF, true, and \iras fields, are all very
similar, indicating that the ITF process does recover the true
peculiar velocity field, as does the \iras gravity field. The
similarity of Figure 4c to 4d demonstrates that the 56 mode
filtering recovers most of the information contained within the
\iras predicted flow field.  The higher resolution velocity
information of the unfiltered \iras flow field has relatively low
amplitude in most regions.

To be slightly more quantitative, we present in Figure 5a the
residual $u_{itf} - u^f_{true}$ as a function of z,
 with points within 60 degrees of
the apex angle toward ($l=270,$ $b=30$) plotted on the right, and
those within 60 degrees of the antiapex plotted on the left.
Note that there is a slight monopole residual, with all distant
points having slightly negative residuals.  This has resulted
from bias in the determination of the Tully-Fisher slope
$\gamma$, which has been fit by the $\chi^2$ minimization
procedure but has led to a biased estimate, $\gamma = -0.0965$,
instead of the correct value $\gamma  = -0.1$.  The bias has been
induced by the thermally ``warm" velocity field of the mock Mark
III catalog, as expected (see ND95 for further details).  If
instead of allowing $\gamma$ to be determined by the $\chi^2$
minimization, we fix it at its correct value, the resulting
$\chi^2$ increases from 1900 to 1905, but the resulting field has
no residual monopole, as shown in Figure 5b.  Figures 2-4, as
shown, were corrected for this bias. We can expect a similar bias
for the real data, since the  scatter in $\sigma_u$ has
approximately the same amplitude as the mock catalog, and we
shall therefore correct the measured TF slope by the same factor
of 3.5\%.  This bias is so minor that none of the results to follow
are affected if we ignore it altogether.

Having corrected for this bias, we next examine maps of
the residual fields.
Figure 5c shows the residuals $u_{itf} - u^f_{iras}$, plotted in the
same manner as Figure 5ab.  Note that the residuals are now somewhat
larger and increase with distance, but show no systematic offsets.
Figures 6a and 6b show the residuals $u_{itf} - u^f_{true}$
and $u_{itf} - u^f_{iras}$ plotted in slices of the
full sky.  Note that the amplitude of Figure 6b is larger than in Figure 6a,
showing that \iras induced errors are a major contributor to the errors
in the procedure. Note again that all the errors are coherent, as
expected since the field is limited to $l=3$; neither the velocity nor
gravity field contain any information at higher resolution.
Comparison of different mock \iras and mock Mark III catalogs
shows similar coherence in the residuals, but largely differing sky
patterns than shown in Figure 6b.
An important point to
emphasize for what follows is that the dipole component of the error in the
mock catalogs is small; the errors appear to be largely dominated by
higher multipoles.  The coherence of the residuals makes it difficult
to judge the adequacy of the fit, but we shall see below when we
consider the statistics of the mode amplitudes that these fits are
consistent with the expected noise.

\bigskip

{\cl {\bf 4.  Comparison of Mark III vs \iras}}

Having set the stage by means of the mock catalogs, we now
consider the actual comparison of the ITF velocity field
from Mark III and the gravity field derived from the \iras 1.2 Jy
catalog.  Although the agreement between the mock Mark III and
the mock \iras catalogs was outstanding, we shall find that the
comparison of the true velocity and gravity fields is
considerably more problematic.

\bigskip

\noindent{\it 4.1 Preparation of the Mark III Database} 

Before constructing the ITF velocity field, it is necessary
to bring the Mark III data to a uniform linewidth and
magnitude system. In its raw form, the catalog consists of six distinct
pieces, each with its own TF relation (Willick \etal
1996a), but by means of direct comparison of overlapping
galaxies, the catalog can be standardized.  We briefly summarize
the procedure here;
complete details are given by Willick
\etal (1996b).  Note that we use only spiral galaxies in the present
analysis.

We assume that all six Mark III samples can be placed on
a ``common system'' by linearly transforming the 
apparent magnitudes and 
and velocity widths, or TF {\it observables.}
For the reasons
discussed by Willick \etal (1995), the Han-Mould (HM) 
cluster sample was used as the template data set, i.e., the
HM data are unchanged in our analysis. 
The remaining samples are then transformed via a hierarchical
overlap-comparison procedure analogous to that described
by Willick \etal (1996a), except that now the TF observables
rather than the TF distances are required to agree in the mean
with those of the same objects appearing in already-transformed
samples. Transformations of the form 
$$
\eta_{\rm common} = a_{0,S}+a_{1,S}\eta_S+a_{2,S}\eta_S^2 \eqno (16)
$$
and
$$
m_{\rm common} = m_S + b_{0,S} + b_{1,S}(m_S-5\log cz) \eqno (17)
$$
are fitted to the overlap data (with one exception, discussed in the next
paragraph), where the subscript ``$S$'' denotes a given sample,
and ``common'' refers to the values of the observables on
the ``common system'' ultimately adopted for our analysis.
The quadratic coefficient $a_{2,S}$ was non-zero only in
the case of the Courteau-Faber (CF) sample, while the coefficient
$b_{1,S},$ which represents a luminosity dependence of galaxy
color, is non-zero only in the case of the $H$-band 
Aar onson \etal (A82) sample.
Values of all coefficients that appear in equations (15) and (16)
are tabulated by Willick \etal (1996b). 

For the Mathewson \etal (1992; MAT) sample, we adopted a
modification to Equation (15). When the MAT
$\eta$'s were initially transformed to the common system,
the fit residuals correlated weakly
with redshift, in the sense that higher redshift MAT
objects appeared to have smaller widths than those of the same
objects in other samples. A possible interpretation is that the
MAT widths, which were measured chiefly in the 21 cm line
at lower redshifts, but primarily optically in the H$\alpha$
line at higher redshifts, are
biased low with increasing redshift. To account for this
possibility, we tried a transformation of the form
$$
\eta_{\rm common} = a_{0,{\rm MAT}}+a_{1,{\rm MAT}}\eta_S
+a_{3,{\rm MAT}}\log(cz). \eqno (18)
$$
This yielded a coefficient $a_{3,{\rm MAT}}=0.065 \pm 0.025,$ which
is small but statistically significant, and a small but significant 
decrease in the scatter of the overlap comparison.
We subsequently found that the ITF velocity fit using the
MAT widths transformed according to (17) produced a
significant reduction in the ITF $\chi^2$ statistic.
We have therefore adopted the redshift-dependent
transformation (17) for the MAT sample 
throughout the analysis presented in this paper;
we discuss possible implications of this further in \S~6.

A final step in the preparation of our modified Mark III
sample involves the rejection of outliers. As has been noted
elsewhere (e.g., Willick \etal 1996a), the TF residuals of
typical complete samples are not strictly Gaussian. 
Rather,
they tend to consist of a gaussian core plus a small
(1--2\%) admixture of 4--5$\sigma$ outliers. 
Such objects are not helpful to velocity analyses, as
they vitiate the assumption of Gaussianity that underlies
almost all statistical methodologies. In our study we
purged such points in the following way. First, we fit
a bulk flow model 
to the entire sample.
We computed the scatter $\sigma_\eta$ from this
fit, and then eliminated the 29 objects which deviated by
more than $3.3\sigma_\eta.$ These rejected galaxies were distributed
throughout the volume of the MAT sample.
When we refitted the bulk
flow models, the scatter was reduced by 10\% as a result
of this small pruning of the sample, while the direction
and amplitude of the bulk flow were essentially unchanged.
Moreover, there were no remaining points which deviated
by $\geq3.3\sigma_\eta$ relative to the reduced scatter.
Thus, our exclusion procedure resulted in a statistically
improved (i.e., more gaussian) sample, without significantly
altering the global properties of the velocity field
the sample produces. 

\bigskip

\noindent {\it 4.2 Maps and Residuals of the Real Sky}

The first stage is to fit the Tully-Fisher data of Mark III to the 56
orthogonal modes.  For 2237 galaxies with $cz <6000 \kms,$ the
$\chi^2$ of the TF regression drops from 3362 to 2616 when
all the modes are included, assuming a scatter of $\sigma_\eta =
0.05$.  The $\eta$ scatter of the different pieces of Mark III is
not identical, but the observed $\eta$ scatter after fitting
the modes is spatially random, so a nonuniformly weighted fit
will be unchanged from the present procedure. 
When a 3.5\% correction is made to unbias the slope of the TF
regression, to $\gamma = -0.112$, the $\chi^2$ increases to 2623, which
is statistically significant, but the resulting flow field is negligibly
changed.  The derived TF slope is $2\sigma$ deviant from the
slope derived by the inverse fits for the Han-Mould sample, $\gamma = -0.1177 
\pm 0.0025$ (Willick \etal 1995, 1996a), but this slope is derived with a 
very different flow model, so we consider the agreement satisfactory.  

 Note that the resulting $\chi^2$
per degree of freedom is 1.15, significantly greater than unity.  The
excess fluctuations are primarily caused by small scale velocity fields
and multi-valued zones 
with coherence scale below the resolution of the mode expansion.  This
is a slightly higher reduced $\chi^2$ than obtained with
the mock catalog, and the increased $chi^2$ could be due to slight
deviations from
Gaussianity in the Tully-Fisher scatter of the Mark III catalog,
an effect we did {\it not} include within the mock catalogs.
Our pruning of outliers from the
Mark III catalog was very modest.

We have examined plots of the distribution of $\delta\eta$ in
redshift and on the sky; all the plots  are uniformly random.  In
Figure 7 we show the correlation function of the $\eta$ residuals
(see e.g. Gorski \etal 1989), before and after the modal
expansion.  This plot demonstrates the residuals to have zero
coherence on scales larger than the resolution scale of the mode
expansion.  It appears as though the mode expansion has done its
job; the expansion of $P_i$ in a set of smooth functions has
greatly reduced the scatter of the inverse TF diagram.  The
randomness of the $\delta\eta$ scatter suggests that the 56 mode
expansion is sufficient for this dataset.  With a larger dataset and/or
more precise distance indicators, a larger set of expansion coefficients
would be justified.

The resulting ITF velocity field of the Mark III galaxies is shown in
slices of redshift space in Figure 8, and projected onto the
Supergalactic plane in Figure 9.  Both of these velocity maps  are in the
Local Group frame.  In the nearby zone, the flow is dominated by infall
to Virgo ($l=284^\circ$, $b=74^\circ$) and Ursa Major
($145^\circ$, $65^\circ$) in the North, and modest 
outflow from Fornax
($237^\circ$, $-54^\circ$) in the south.  In the middle redshift zone,
the dipole pattern persists, and is more pronounced. This is composed
partially by reflex dipole of the motion of the LG, plus backside
infall from behind the Virgo supercluster. The Hydra cluster seen at
($270^\circ$, $26^\circ$) is observed to be falling toward us, while the
foreground of the Centaurus region ($310^\circ$, $20^\circ$) is moving
away from us.  In the most distant slice of redshift space, the dipole
pattern is further enhanced, with the entire Southern galactic sky,
including the Perseus-Pisces region ($150^\circ$, $-15^\circ$) and the
Pavo-Indus-Telescopium region ($320^\circ$, $-15^\circ$) flowing away
from us, the Centaurus cluster flowing away rather substantially, but
the background of Hydra and the Northern galactic cap flowing
toward us.  All of this is a combination of the reflex of the motion of
the LG, plus the motion of the individual regions.  Note the very
strong shear in the observed ITF field along the boundary $l=300,$ $b>0$,
which divides the Hydra and Centaurus regions, and amounts to a change
of radial peculiar velocity of approximately $900 \kms$ in a transverse
interval of $4100 \kms,$ or 20\% of a Hubble flow differential motion
over a substantial volume!  This is quite a major gradient,  one that
appears to have been confirmed by distance estimates of elliptical
galaxies in Centaurus (Lynden-Bell \etal 1988, Dressler 1994). In fact
the true gradient is limited by the angular resolution of our modal
expansion, since the reported outflow of the Centaurus clusters is
$500$--$2000 \kms$ in the LG frame, whereas the ITF measured flows for the
entire Centaurus cluster are in the range $350$--$450 \kms$ in the LG
frame.

For comparison, we show in Figure 10 the \iras predicted field for
$\beta = 0.4$ on shells of redshift, and as projected onto the
Supergalactic plane in Figure 11.  Note that the Mark III galaxies have
an \iras  predicted mean Hubble flow of 2.9\%, which was removed in the
process of smoothing the \iras predicted field by the mode expansion.
Residual fields are shown in Figures 12-14 for $\beta =
0.4$ and $0.7$. 
Note the similarities and differences between these Figures and 
Figures 5-6 for the mock catalog. 
It is important to remember that the ITF
velocity and \iras gravity fields have been processed completely
independently of each other. To a first approximation, the agreement
between these Figures is a spectacular and remarkable confirmation of
the consistency between the gravity and velocity fields of our local
Universe.  The \iras galaxies must have some close relationship to the
underlying mass distribution. The qualitative alignment of the fields
is consistent with the large scale structure having been formed by the
process of gravitational instability.

On the other hand, the residual fields, for any value of $\beta$, are
not nearly as clean as seen in the mock catalogs, and are always
dominated by strong dipole patterns that increase linearly with
redshift. This is especially pronounced for the higher values of
$\beta$ and is particularly clear in Figure 14.  The \iras maps
never reproduce the strong shear seen in the ITF maps between the
Hydra and Centaurus regions. They never reproduce the strong
infall measured by ITF for the $l=180^\circ,$ $b=45^\circ$,
$cz=5000 \kms$ region. For $\beta \le 0.3$  (not shown), the
local supercluster flow is much too small, and the flow toward
Perseus-Pisces is too small.  For $\beta= 0.4$, the residual
field in the local zone is still completely coherent, but the
flow toward Perseus-Pisces and the average of Hydra and Centaurus
is approximately correct.  For $\beta = 0.7$ the relative motion
toward Virgo  matches the ITF measurement ($-340 \kms$ in the LG
frame) but the flow away from Perseus-Pisces and toward the Great
Attractor complex is now too large.  

\bigskip

\noindent{\it 4.3 The Dipole Residual on Shells}

The residual field between the ITF and \iras fields is clearly
 dominated by the dipole component.  ND94 emphasize
that the dipole component of the radial peculiar velocity field on a shell, 
relative to the LG, 
is generated entirely by the mass distribution within that shell;
the exterior mass distribution affects only quadrupole and 
higher $l$ perturbations.  Thus it is critical to seek 
agreement of the dipole components of the fields
on modest scales, $cz<6000 \kms,$ where the data are best.  

Although the large spatial inhomogeneity of the Mark III data strongly
mixes the modes together, it is possible to extract the dipole
component of the field.  Consider binning the Mark III data into shells
of redshift $1000 \kms$ thick.  Then compute the three components of a
dipole $v_D$ by summing the radial peculiar velocity $v_j$ 
over the $j$ galaxies in the shell:
$$ v_D = {\sum_{j} v_j w_j {\rm cos}(\theta) \over 
\sum_{j} w_j {\rm cos}^2(\theta) } ~,
\eqno (19) $$
where $\theta$ is the angle from the three unit vectors ${\bf
\hat x, \hat y, \hat z }$. By using a 
weight function $w_j$ given by $(\delta\theta_4)^{-2}$, where
$\delta\theta_4$ is the angle of galaxy $j$ to its fourth nearest
neighbor on the given shell, the computation of $v_D$ is approximately
equal area weighted, as is appropriate for comparison to the
original functional expansion.

An example is shown in Figure 15, where we plot the ITF and \iras (for
$\beta = 0.6$) derived dipoles in six shells within $6000 \kms.$  The
open and closed symbols are derived from the irregularly spaced Mark
III galaxies, while the solid lines are original \iras dipole
components, for $\beta = 0.6$, based on the full 1.2 Jy sample.  Note
that the \iras derived points, drawn from a less than optimal
sky distribution, still match the
smooth curve very well, demonstrating that recovery of the dipole from
the irregularly spaced Mark III sample is not a problem.  The open squares
showing the ITF dipole are consistent with the \iras dipole in the Z
component, and the total amplitude is a good match as well.  But the X
and Y components are seriously discrepant. The X component is
consistent for the first three shells, but then deviates substantially,
while the Y component is discrepant for all shells; \iras predicts a
substantial reflex dipole in the Y component that should grow with
redshift, but little is detected in the ITF data.  For those who have
followed this subject for some time, this discrepancy is not new (see
e.g. the comparison of Faber and Burstein (1988) with the discussion by
Strauss and Davis (1988) and Yahil (1988)).  The \iras field expects to find
substantial ``backside infall" at the Centaurus region and beyond, 
but such a signature is very weak in the Mark III data.
We 
emphasize once again that the comparison under consideration here is for
spiral galaxies only; the discrepancy is observed for both spiral
and elliptical galaxies.

\bigskip
{\cl {\bf 5. Likelihood Analysis of Modal Comparison}}

\noindent{\it 5.1 The scatter of ITF and \iras coefficients}

 Given the measurement of the coefficients $\alpha^j$ 
derived from the Mark III and \iras\  fields, 
how can we best compare them?  In the
discussion above, we compared the fields galaxy by galaxy, and by
means of the area weighted dipole on shells. 
The sky maps are 
appropriate for visualization; however the fields as seen in space are
highly correlated because of the strong covariance of the errors in the
ITF velocities as well as in the \iras gravity.  
Since the number of
modes which describe the fields is significantly less than the number
of galaxies in the Tully-Fisher sample, a galaxy by galaxy comparison
requires inversion of an unduly large covariance matrix.  Therefore we
work directly with the recovered mode coefficients from the ITF and the
\iras gravity in order to assess the quantitative 
agreement between the velocity and
gravity fields.

Figure 16 plots the coefficients   $\alpha_{tf}$ versus  $\alpha_{iras}$
for the real data, using $\beta= 0.6$ for the \iras coefficients. 
Also shown are the coefficients for one realization of the mock
catalog; in this case we can also plot the coefficients $\alpha_{tf}$ versus
$\alpha_{true}$.  Note in the mock case that the coefficients track the
diagonal line extremely well, consistent with $\beta =1$. As expected, the
noise is increased for the ITF versus \iras comparison. Different realizations
of the mock catalog all have consistent behavior in their mode-mode
scatter diagrams.

\noindent{\it 5.2 The error matrix on ITF and \iras coefficients}

As is shown in ND, the orthonormality condition leads to
the following simple form for the error matrix of the ITF coefficients,
namely,
$$ \langle \delta \alpha^{j}_{tf} ~ \delta \alpha^{\jp}_{tf} \rangle = 
\left({\sigma_\eta \over s}\right)^2 \delta^K_{j\jp}  ~. \eqno (20) 
$$
In Figure (16), the plot of $\alpha_{tf}$ versus $\alpha_{true}$
for the mock catalog has noise that is completely consistent with equation
(20), as expected. 

 For the Mark III catalog, six separate 
samples, with
differing photometric bandpasses and velocity width methods, have been pieced
together in the manner described above.  
The offsets in the magnitudes
and linewidths for this adjustment have uncertainty in the mean, and
this will translate into  a non-diagonal
contribution to the covariance matrix 
$< \delta \alpha^{j}_{tf} ~ \delta \alpha^{\jp}_{tf} >$.  We estimated
this contribution by generating 20 realizations of the Mark III catalog
with gaussian errors in the zero point of the different pieces appropriate
for each subsection.  We then directly estimated the covariance matrix
generated by these offsets, and included them in the analysis below.
In fact, the errors in the zero points are small and make no difference
to the error budget.  

The situation for the modes of the \iras predicted velocity 
field is not so simple, however.  
The error matrix $ < \delta \alpha^{j}_{iras} ~ \delta \alpha^{\jp}_{iras} >$
has to be contructed from the covariance matrix of the \iras predicted
velocities of galaxies. Using Equation 
(15),  and linearizing the errors in the velocities
about the predicted values yields,
$$M^{j\jp} \equiv < \delta \alpha^j_{iras} ~ \delta \alpha^{\jp}_{iras} > = 
\left({5 \over 2.3}\right)^2 \sum_{ii'} 
{< \delta u_i \delta u_{i'} > \over z_i z_{i'} } 
 {1 \over (1 - (u_i / z_i)) } {1 \over (1 - (u_{i'} / z_{i'})) } 
\Ft^{j}_i\Ft^{j'}_{i'} \eqno (21) 
$$
where $< \delta u_i \delta u_{i'} >$ is the covariance matrix of the
\iras  predicted velocities of galaxies $j$ and $j'$ after subtracting the
Hubble-like flow as described in {\it 2.2.2}.  Note that even
if this velocity covariance matrix is diagonal with equal elements, 
the covariance matrix of the coefficients will not be
diagonal, as the modes $\Ft$ are orthogonal with unit weighting of each
point, whereas the sum above is  weighted approximately as
$z_i^{-2}$.

The  \iras velocity field is a very nonlocal quantity, effectively
computed by counting the vector sum of all \iras galaxies around a
given point, weighted by the inverse of the square of the 
real space separation
and by the inverse of the selection function.  This sum is
computed over a discrete set of points in a finite catalog, and
therefore must suffer from Poisson noise 
and from variance induced by
the assumption that the mass per \iras galaxy is assumed constant for each 
redshift shell (c.f. Strauss \etal 1992).
In addition, the computed velocities assume linear theory, yet there
exist high density zones within the galaxy distribution for which
linear theory is known  to be inaccurate and for which the deviations
are likely to be coherent (as in triple-valued zones or 
zones which have shell crossed).  There
are incoherent small-scale nonlinear velocities 
unresolved by the \iras procedure
which add noise as well.  Furthermore, the \iras predicted velocities
are sensitive to the assumed transformation from heliocentric to 
LG frame velocities, which might be in error by perhaps 50-100
$\kms.$  All of these error sources contribute to errors in the \iras\
reconstructed velocities, and most of these errors are coherent.

We can express the \iras\  velocity covariance as a sum over these terms,
writing 
$$
  <\delta u_i \delta u_{j} > = 
 C_{SN}(i,j) + C_{NL}(i,j) + \sigma_{LG}^2 \cos(\theta_{ij}) \eqno (22)
$$
where $C_{SN}(i,j)$ is the covariance in the
predicted peculiar velocities due to the Poisson noise, $C_{NL}$ is a
term to describe the deviations from linear theory,   and $\sigma_{LG}$
is the uncertainty in the transformation to LG frame. We shall
set $\sigma_{LG} = 100 \kms.$ Note that this
term alone generates complete covariance of all \iras predicted
peculiar velocities.

The Poisson noise  term $C_{SN}(i,j)$ can easily be derived from
a distribution of points in $real$ space (Yahil \etal 1991)
where, according to linear theory,  the peculiar velocity field
is related to the distribution of galaxies by the Poisson
equation, and the variance $C_{SN}$ would scale as $\beta^2$.
However, the galaxies are observed in redshift space and the
error in the predicted velocity field must be computed according
to the relation (4), for which the scaling with $\beta$ is not
trivial.  An analytical expression for $C_{SN}$ for the
velocities from (4) is cumbersome and numerically expensive to
evaluate.  Thus we choose to compute the covariance matrix
$C_{SN}$ by generating 18 bootstrap realizations of the observed
1.2 Jy \iras galaxy distribution by replacing each observed
galaxy with a number of points drawn from a Poisson distribution
with mean of unity. Then we compute the velocity fields from
these realizations by the same algorithm used in the derivation
of the velocity field from the observed galaxy distribution.  For
galaxy $i$ of the Mark III sample, we tabulate the differences,
$\delta u_i$, between the velocity obtained from each of 18
bootstrap realizations and the velocity as predicted from the
actual distribution of galaxies, and evaluate the covariance
matrix $C_{SN}$ by averaging the product $\delta u_i\delta u_j$
from all the bootstrap realizations.  This process is computed
for several different values of $\beta$.

The term $C_{NL}(i,j)$ describing nonlinear effects can be modelled as 
$$ 
C_{NL}(i,j) = \sigma_u^2 \delta_K^{ij} + \left((s_h(i) + s_h(j)\right)^2 
{\rm exp}\left(- {|\bfr_i - \bfr_j|^2 \over 2 \sigma_{coh}^2 } \right)  
~, \eqno (23) 
$$
where particles have an assumed incoherent small scale 1-D velocity dispersion 
$\sigma_u$, but
in regions of high shear can also display coherent errors, with coherence
length $\sigma_{coh}$ which we set equal to $300 \kms.$  Miller \etal (1996)
have directly measured $\sigma_u$ for the \iras 1.2 Jy survey, and
derive a best value $\sigma_u = 137 \kms.$  Willick \etal (1996c)
find a best value of $\sigma_u \approx 150 \kms$ for the Mark III
galaxies using the VELMOD analysis 
(i.e. ``Method II$^+$" as described by Strauss
and Willick (1995)). The Mark III and \iras galaxies need
not have the same thermal dispersion, but the agreement is 
comforting.  We shall assume a value $\sigma_u = 150 \kms.$ 
The coherent error
we assume is proportional to the square of the average shear in the
\iras derived velocity field,  
$s_h(i) \equiv 100~{\rm min}(1,|d v_p/dz|) \kms,$
which is 
an approximately linear function of $\beta$.  This term is designed to
deweight points in regions of probable shell crossing, where the \iras
predicted shear is large. 

\noindent{\it 5.3 Pseudo $\chi^2$ analysis}

Our basic assumption is that, for the true $\beta$, the ITF and the \iras 
velocity fields are noisy realizations of the true underlying
velocity field. Let $\aj$ be the coefficients obtained from  the modal 
expansion of the true underlying field. 
Each estimated coefficient is derived from a sum over a large set
of numbers, so by the central limit theorem we expect each estimated
coefficent to have a normal distribution, even if the underlying velocity
field or Tully-Fisher data has nongaussian scatter.
Since the Mark III and the \iras data are independent we write
the probability for measuring $\aj_{iras}$ and $\aj_{tf}$ as,
$$
P\left[\aj_{iras},\aj_{tf}|\aj\right]=A 
\exp \left [-{1\over 2}\left(\chi^2_{tf}+\chi^2_{iras}\right)\right]
\eqno (24)
$$
where we have defined,
$$
\chi^2_{tf}=\sum_j{{\left(\aj-\aj_{tf}\right)^2}\over {\sigma^2_{tf}}}
$$
and,
$$
\chi^2_{iras}=\sum_{j,j'}(\aj-\aj_{iras})({\rm M}^{-1}_{jj'})
(\ajp-\ajp_{iras}).
$$

Note that $\chi^2_{tf}$ does not depend on $\beta$.
Given $\beta$, we seek the true coefficients, $\aj$, and $\beta$,
by maximizing the
probability (24). Therefore our best estimates are obtained by
solving,
$$\partial P/\partial \aj=0 , 
\eqno (25)
$$
and,
 $$
\partial P /\partial \beta=0 .
\eqno (26)
$$
Since the normalization
of the probability (24) does not depend on the coefficients $\aj$,
the condition (25) implies that the parameters $\aj$ which
maximize $P$ for a given $\beta$ renders  a mimimum in the
total 
$\chi^2\equiv \chi^2_{iras}+\chi^2_{tf}$.
However, since the error matrix ${\rm M}$ and hence the normalization
factor $A$ strongly depend on
$\beta$, the maximum of $P$ with respect
to $\beta$ does not necessary coincide with the minimum of $\chi^2$. 
Though, at a given $\beta$, the condition (25) yields a set of linear equations
for $\aj$ which can be solved using standard methods, the
condition (26) leads, in general, to a complicated nonlinear
algebraic equation for $\beta$. 
Therefore, a simultaneous estimation of the coefficients
$\aj$ and $\beta$ is a problem of nonlinear minimization. Fortunately,
in this particular case, it is easy to see that solving (25) and (26)
simultaneously can be achieved by the following scheme: 
solve (25) for $\aj$ at a given $\beta$ and compute the corresponding
value of $P(\beta)$. The estimate for the true $\beta$ is the
value which renders $P(\beta)$ a maximum.
The condition (25) yields, 
$$
\vaj \cdot \left({\bf I}\sigma^{-2}_{tf}+{\bf M}^{-1}\right)=
\vaj_{tf}\cdot {\bf I}\sigma^{-2}_{tf}+\vaj_{iras} \cdot {\bf
M}^{-1} .
\eqno (27)
$$
By  substituting (27) for $\aj$ in the total $\chi^2$, we have,
$$
\chi^2|_{\partial \chi^2 /\partial \aj=0} 
\Rightarrow \chi^2_{tf} + \chi^2_{iras} = 
\sum_{j,j'}\left(\aj_{tf}-\aj_{iras}\right)
({\rm I}\sigma^2_{tf}+{\rm M})^{-1}_{jj'}\left(\ajp_{tf}-\ajp_{iras}\right) 
, \eqno (28)
$$
which reduces to the pseudo $\chi^2$ suggested by Press \etal
(1992) for problems with errors in both the $x$ and $y$ directions.
Note that the difference between the values of $\beta$ which
minimize the $\chi^2(\beta)$ and maximize the probability $P(\beta)$
must be consistent with the errors in their estimates 
provided that the ITF and \iras coefficients 
differ only because of the noise in their measurements. 

Comparing the Tully-Fisher coefficients $\aj_{tf}$ to \iras
derived coefficients $\aj_{iras}$ for the mock catalog gives
$\chi^2 = 53.5$ for 56 degrees of freedom.  Thus the fit of the
coefficients of the mock catalogs is entirely consistent with the
expected noise, and this consistency is seen in all the
realizations of the mock catalog.  When we apply equation (22) to
the mock catalog, comparing  the Tully Fisher coefficients
$\aj_{tf}$ to the {\it true} coefficients $\aj_{true}$, while
holding $M_{ij}$ nonzero, we obtain $\chi^2= 30.2$ for 56 degrees
of freedom. This low $\chi^2$ results from inappropriate
inclusion of the \iras like covariance; the approximate halving
of the $\chi^2$ demonstrates that the errors introduced by the
imprecision of the \iras estimates are similar in magnitude to
those of the ITF estimate.


The results of applying this same procedure to the real data is
given in Table 1, where we list $\chi^2$ versus $\beta$.  For
each value of $\beta$ we first computed $M^{jj'}$ (equation 16)
and then evaluated equation (23).  The table shows that the
minimum $\chi^2 = 100$ (55 d.o.f.)  at $\beta = 0.5$.  Given that
$\chi^2$ for $\nu$ degrees of freedom has an expected mean of
$\nu$ and a variance of $2\nu$, the best $\chi^2$ is $4.3 \sigma$
deviant from its expected value.  The fit is statistically very
improbable and can be rejected.  That is, although the mock
catalogs demonstrate that the procedure is capable of producing
very acceptable $\chi^2$ results, the actual comparison of Mark
III to \iras leads to the conclusion that the fields are statistically
inconsistent with each other.  The details of the best $\beta$
and the level of inconsistency are weak functions of how the
catalogs are limited; we would derive a higher $\beta$ if we
limited the catalogs to $cz < 4000 \kms,$ but the fit would
remain unacceptable.  

Because the residual field is dominated by a strong dipole and 
because we believe we understand the error matrix, 
it is not permissible to simply double the covariance matrix to
bring the $\chi^2$ values to reasonable levels.  Thus we urge
extreme caution in concluding that $\beta=0.5 \pm 0.1$ is in any
sense a ``best fit" solution.

\bigskip\bigskip
{\cl {\bf 6. Discussion}}

\bigskip
\noindent{\it 6.1 Alternative Methods and Results}

The ITF analysis is only one of several, independent methods of comparing
the Tully-Fisher catalogs with the redshift survey catalogs.  Given the
difficulty of the comparison and
the subtlety of the error analysis, it is worthwhile to investigate
alternative approaches and to expect consistent results.  Only then will 
we have confidence in the derived value of $\beta$.

The great advantage of the ITF scheme is that it allows the comparison of
the velocity and gravity field mode by mode, with identical resolution
to each field. The Monte-Carlo simulations with realistic
mock catalogs confirm that the bias in the extracted velocity field
is negligible. The method allows for the gradient in radial resolution of the
two fields, so it is capable of extracting the maximum useful
information for each field.  One can rank order the modes by
their significance, or contribution to $\chi^2$, keeping only
the most significant modes, so as to provide a very compact
description of the field. This is an example of a Karhunen-Lo\`eve
transformation (Therrien 1992). Recent applications of this
technique to astrophysics include the CMBR fluctuations of the
COBE data (Gorski 1994, Bunn 1995, Bond 1995), the power spectral
analysis of fluctuations in redshift surveys (Vogeley  and Szalay 1996), and
the spectral classification of galaxies (Connolly \etal 1995).

The ``VELMOD'' approach of
Willick \etal (1996c; cf.\ the
preliminary discussion presented by Strauss \&
Willick 1995, \S~8.1.3) is in several respects complementary
to the analysis presented here.
\velmod also makes a comparison
of the Mark III
TF data and the gravity field derived from the \iras 1.2Jy redshift survey.
It differs from the ITF analysis in explicitly accounting
for the small-scale velocity dispersion $\sigma_u,$ which it
treats as a free parameter, and in allowing for the possibility of
multivalued redshift-distance diagrams; the ITF reconstruction
implicitly assumes a unique redshift-distance mapping.
Implementation of \velmod is done
using grid-based, real-space \iras velocity reconstructions, as
opposed to the noniterative redshift-space \iras velocity
fields used here. Thus, \velmod and ITF do not use identical
gravity fields, although tests have shown them to be very similar.

There are several advantages to \velmod. First, 
in accounting for small-scale velocities and arbitrary
redshift-distance diagrams, it determines the probabilities
of individual data points without some of the 
approximations used here, and thus provides a reliable
maximum likelihood statistic. It does not
require that the input TF catalog be placed on a common
system, but instead treats the individual sample TF relations
as free parameters. This makes it free of the
uncertainties induced by the overlap-comparison procedure
described above (\S~4.1). Finally, \velmod is particularly 
well-suited to analysis of the nonlinear regime, which holds
the promise of breaking the $\Omega$-$b$ degeneracy. However,
\velmod has an important weakness from which
neither ITF nor POTENT suffers: it does not permit the construction
of a velocity field from the TF data alone; instead, it assumes
that the \iras velocity field, for some value of $\beta$ (and $b$), is
a good fit to the data, and selects the best $\beta$ by maximizing
likelihood. Yet as we have shown in this paper, 
the \iras velocity field does not fit the data well
throughout the sampled volume. In such a circumstance,
the meaning of the maximum likelihood statistic is unclear.
In addition, \velmod is
not well suited to a proper analysis of covariant 
velocity field prediction errors, such as we have
carried out here.
 
A final distinction derives from the depth of the respective analyses.
\velmod is rather cpu-intensive and, moreover, differs
meaningfully from analyses such as ITF only when $\sigma_u$ is 
a substantial fraction of the TF distance error. 
The analysis of Willick \etal (1996c) is
thus restricted to redshifts $cz\leq 3000 \kms.$ In contrast,
we have
extended our computationally more efficient analysis to
twice that distance. As noted above, our low indicated value of 
$\beta=0.4-0.6$ 
is in part a consequence of discrepancies between the ITF
velocity field and the \iras gravity field that are manifest
principally at distances $\simgt 4000 \kms.$ 
Willick
\etal (1996c) analyze their likelihood fit residuals and find
them to be nearly incoherent across the sky, which reflects the
fairly good agreement between \iras and the TF data within $\sim 3000 \kms.$
Preliminary results seem consistent with our ``best" values of 
$\beta$. 

  Both the POTENT and ITF methods have 
been validated by extensive testing with mock catalogs, so it seems
reasonable that they should yield consistent results.  
We here list several several relative advantages and disadvantages 
between the ITF method and the POTENT/IRAS comparison.

\item{(1)} POTENT provides a measure of the divergence of the velocity
field which can be compared to the directly observed galaxy
density field.  Both fields are local, an obvious advantage.  The
ITF field must be compared to the non-local \iras derived
velocity field. These fields are both susceptible to systematic
errors within the data.

\item{(2)} POTENT required a large, uniform smoothing to properly
define the smoothed velocity and velocity potential.  Dekel \etal
(1993) have been using a gaussian smoothing of width $\sigma = 1200 \kms,$
sufficient to filter out the infall of the Local Group toward
the Virgo cluster.  
The ITF method can work with much smaller smoothing, and the
smoothing length can be dependent on redshift.  This is
quite advantageous, since the statistical precision of both the
velocity and gravity field are decreasing functions of distance.  
On the other hand, the ITF method can readily accomodate uniform smoothing,
if that is desired.

\item{(3)} The POTENT 
analysis is dependent on a calibrated forward Tully-Fisher
relation (i.e. $M= M(\eta)$, where $M$ is the absolute magnitude
and $\eta \equiv \log(\Delta v)$ is the velocity width parameter), and the
analysis is sensitive to Malmquist bias (see Strauss and Willick 1995
for details).  The ITF analysis uses an {\it inverse\/} TF
relation (i.e. $\eta = \eta(M)$); 
the parameters of this relation can be fitted at the same
time as the mode coefficients.
As long as the galaxy selection is independent of
linewidth, this method does not suffer from Malmquist bias.  
The ITF method is biased in the presence of a thermally hot velocity field,
where multistreaming is very pronounced; however, given the estimated
amplitude of the small scale peculiar velocity field of galaxies, this
bias is quite modest, as described above (see ND95 for details).  

\item{(4)} The
ITF method is simply a redshift space
 smoothing algorithm for the Tully-Fisher data that enables one to get
a measure of the velocity field without 
binning the
data.  It  is not intended for the construction of full-sky velocity
maps, and such maps are not needed for comparison to the gravity
field.  The method compares the two fields only where there are data.
By avoiding binning of the data, it eliminates the ``sampling
gradient bias" that is a serious problem for POTENT.  
Both the ITF and \iras analyses are here carried out in observed redshift
space, whereas most previous analyses, including POTENT,
 have been done in real
space.  Since the inversion to real space is not unique in the multi-valued
zones, attempting to perform the ITF analysis in real space would
present additional very serious complications.  However,
with sufficiently precise distance indicators, working in redshift space  
will not take full advantage of the data, since multivalued 
zones could be resolved (e.g. Strauss and Willick 1995).

\item{(5)} By filtering the density and velocity fields with
identical functions, one can be confident the fields have identical
resolution. In the POTENT analysis, assessing the relative resolution
of the galaxy density field compared to the inferred mass density field
is not straightforward. Previous velocity-velocity comparisons have been
especially compromised by the inability to filter the velocity field. 
Although we have here chosen a very general
set of functions for the mode expansion, if desired, it would be 
straightforward to supplement this expansion with additional functions to 
describe known
features, such as Virgocentric infall.  This would be an appropriate route
for  datasets smaller  or more concentrated in selected  
directions than Mark III.

\item{(6)} The full covariance matrix of the coefficients of the ITF and \iras
fields can be computed, and because the matrix is modest in size, it can
be readily inverted.  This is essential for a proper statistical
assessment of the fit. 
POTENT compares velocity/dentiy fields computed 
on a grid, and the error in $\beta$ must be estimated using sophisticated 
Monte Carlos methods.  
All previous point by point velocity-velocity
comparisons have failed to include this essential covariance in the
error matrix of the gravity field.

\bigskip

\noindent{\it 6.2: Why Don't the Fields Agree?}
 
The ITF method has led to an unambiguous measure of the Tully-Fisher velocity
field, one we feel is correct within the limited resolution of our chosen
expansion.  The biases generated by small-scale thermal velocities are modest. 
There is undoubtedly additional information that could be gleaned
from the Mark III database, as our basis function of modes is not necessarily
optimal for the derivation of the divergence of the flows, the critical 
information extracted in the POTENT analysis.  The tests with the mock
catalog validate our approach, and demonstrate that comparison of the Mark III
data with the \iras derived field should yield consistent fields.  The
error matrix of both the Mark III and \iras derived mode coefficients is, we
believe, well understood.  Yet the comparison of the modes conclusively 
demonstrates that they are not consistent within their errors 
at the 4-$\sigma$ level.  We have already used a very generous Tully-Fisher
error, $\sigma_\eta = 0.05$ (corresponding to $\sigma_M \approx
0.45$). Increasing this error will reduce the formal $\chi^2$ but
will not remove the coherent residuals.

The nature of the inconsistency is dependent on the depth of the sample
under consideration.  To fit the Local Supercluster will require
$\beta \simgt 0.7$, but the large dipole residual at large scales demands
a smaller $\beta$.  Thus it is not surprising that the ``best" $\beta$ for
our analysis, $\beta = 0.4-6$, is slightly lower than suggested by \velmod,
$\beta=0.57,$
as that analysis is confined to galaxies with $cz< 3000 \kms.$

The 4-$\sigma$ discrepancy between the \iras derived velocity field and
the Mark III velocity field suggests
that the problem is not due to random errors in either field, but to
systematic problems in one or both datasets.  There are many possible
sources of systematic errors, and we here list and discuss the most
obvious among them.  It is very likely that some combination of these
errors is responsible for the statistically poor correspondence of the
fields, and until we can deal with the major problems, all derivations
of $\beta$ from flow analysis must be considered suspect.

\item{(1.)} {\it The \iras field is missing the early type
galaxies, so it ignores cluster centers.}  Careful comparison of
the counts of \iras galaxies to optically selected galaxies
indeed shows an underrepresenation of the galaxies within the
centers of clusters (e.g. Virgo's core is underrepresented by a
factor of 2).  However the cores of clusters do not dominate the
mass of the superclusters (recall that the cluster mass typically
diverges as radius R), and the \iras galaxies appear to be good
tracers of the general field population (Strauss \etal 1992).
``Boosting" known cluster centers to compensate for the missing
early type galaxies does not lead to a better fit between the
gravity and velocity fields, but it will reduce the best estimate
of $\beta$ by a factor of $\approx 1.3$.


\item{(2.)} {\it The \iras field does not have whole sky coverage
and is missing a prominent cluster in the Zone of Avoidance}.
Recently Kraan-Korteweg \etal (1996) have reported on a Coma-like
cluster, A3627, with estimated mass $2.5
\times 10^{15} h^{-1}~\Msun$ at position l=325$^\circ$,
b=$-7^\circ$, $cz_{LG}
\approx 4700 \kms.$  
Although the \iras survey does include galaxies to $|b| >
5^\circ$, and the observed galaxy density is interpolated through
the Galactic plane, a cluster of entirely early type galaxies
would have been severely undercounted.  The \iras catalog has an
overdensity of galaxies at the position of A3627, but it is not
sufficient to comprise a large Coma-like cluster, and it is not
by any means the most massive \iras cluster in this vicinity.
Could this possibly explain the strong shear observed in the ITF
analysis between the Hydra and Centaurus regions?  Suppose that
the \iras survey has given no mass to A3627.  Then for the
parameters above, the neglect of A3627 has led to an error of the
peculiar motion of the LG of $\approx 50
\Omega^{-.4} \kms.$  We can readily compute the missing
differential motion between the Centaurus clusters at (l,b) =
(302$^\circ$, 22$^\circ$) and the LG caused by A3627.  If
Centaurus is at distance of $\approx 3800 \kms,$ as suggested by
the \iras field, then its acceleration toward A3627 is in the
transverse direction.  If the distance of the Centaurus clusters
is lowered to $2500 \kms,$ as suggested by several distance
estimates (Lynden-Bell \etal 1988; Dressler 1994), 
then the differential radial acceleration between
Centaurus and the LG induced by A3627 is only $4 \Omega^{-.4}
\kms.$  The differential influence of A3627 on Hydra (at (l, b,
cz) = (275$^\circ$, 20$^\circ$, 4500 km/s)) and the LG is
approximately $75\Omega^{-0.4} \kms.$  These results scale
linearly with the mass of A3627.  Thus as one increases the mass
of A3627, the residuals toward Hydra become more negative whereas
the residuals toward Centaurus are changed by a negligible
amount.  This is the wrong sense, since the residuals toward
Centaurus are far worse than those toward Hydra.  Furthermore,
the effect of a missing cluster like A3627 is modest compared to
the residuals to be explained.  Examination of the \iras density
field at this redshift demonstrates that the total mass
overdensity in the Hydra-Centaurus-Pavo-Indus complex greatly
exceeds that of any missing cluster center, and that the addition
of the core of A3627 makes no difference to the overall density
field.

\item{(3.)} {\it The \iras field has poor convergence to the CMB
dipole either because of dilute sampling noise at large distance
or because of an improper transformation to the Local Group
frame.} The ITF method is sensitive to the choice of the LG
frame.  The Mark III data, when analysed by the POTENT algorithm,
exhibits a bulk motion of $250$--$300 \kms$ that
is not observed in the \iras derived gravity field (Dekel 1995).
If this is due to an incorrect choice of the LG frame, then the
effect is in the direction of the nearby Maffei/Dwingeloo group
of galaxies (Kraan-Korteweg \etal 1994), which are highly
obscured and are not included in the
\iras gravity field.  If they are the source of
this anomalous motion, then they should be blueshifted by our infall toward
them.  However they are observed to have redshifts, not blueshifts, and so
must move coherently with the Milky Way.  Beyond the local galaxies in this
direction is a void, which should repel us, not attract us. It is
difficult to understand how the LG could have a local source of
acceleration leading to a velocity as large as that induced by
the Virgo supercluster, without leaving any signature in the
local galaxy distribution.  The local flow field has only modest
shear, and a large error in the LG frame definition is unlikely.
This is apparent in Figures 8, 14, and 15; for shells within
$2000 \kms,$ the dipole residual between the ITF and \iras fields
is very modest.  An error in the LG frame would lead to dipole
residuals {\it constant} with redshift, whereas the observed
dipole residual grows approximately linearly with distance
(Figures 14 and 15).  The local flow is rather quiet (Peebles 1992; Schlegel
\etal 1994) and shows no
evidence of the very large local shear that would result if the
LG frame were offset by several hundred $\kms.$

\item{} As a test of the redefinition of the LG, we have generated \iras fields
and ITF fields with the LG frame offset toward Maffei by $200 \kms.$ Such
an offset leads to smaller residuals toward Perseus-Pisces for high values
of $\beta$, compared to Figure 14, but instead there are coherent
errors toward Hydra and other nearby regions, as expected. We were unable to
achieve a better fit by adjusting the LG frame.

\item{} The diluteness of the \iras density field could lead to substantial
errors in the predicted $u_{iras}$ field, especially from matter at
redshifts $cz> 6000 \kms$ (Davis \etal 1991). The mock catalogs demonstrate the
degree of expected error.  However, errors in the density field of external
matter will not affect the $l=1$ residuals in the LG frame.  These dipole
residuals, if real, are indicative of a mass field not traced by the \iras
galaxies.  Since we expect linear theory to apply, we can simply estimate
the unseen mass density by taking the divergence of the residual dipole
field.
Making allowance for the transverse fields by assuming potential flow,
and noting that the residual field scales approximately linearly with
redshift, we find a residual mass field $\delta_\rho \propto {\rm cos}\theta$,
where $\theta$ is the angle from the apex direction.  This is a very strange
density field, one that is constant in redshift but tilted toward a particular
direction.  Thus the residual dipole field cannot be considered
physically reasonable; it must be indicative of
some systematic discrepancy. 

\item{} The reason our results are not consistent with the
high $\beta$ values derived from the density/density comparisons 
(Dekel \etal 1993) using the 
POTENT algorithm is clear from Figure 14.  For large values of
$\beta$ the dipole residual  growing linearly with $z$ is very pronounced. It
cannot be removed by a transformation of the LG
center, nor can it be explained by an external mass field.  It can only be
reduced by elimination  of some unknown systematic error in one of the catalogs, 
or the reduction of $\beta$.  The POTENT algorithm, by taking
the derivatives of the Mark III velocity field, is less
sensitive to this large scale dipole, but it is more sensitive to higher
spatial frequency components of the flow field.  This very different
weighting of the data, combined with the poor quality of the fit of
\iras to Mark III for any value of $\beta$, is sufficient to explain why 
it is not unreasonable for 
the ITF/VELMOD algorithms to yield ``best" $\beta$ values that are
inconsistent with the best estimates derived from POTENT using the same data.
One could argue that the difference in the derived values of $\beta$ is
the result of a strong scale dependence of the  biasing relation. 
However, as shown
by Kauffman, Nusser and Steinmetz (1996), all
physically reasonable bias mechanisms
for hierarchical models of structure formation are well approximated as linear,
scale independent biasing for scales larger than a few Mpc.


\item{(4.)} {\it The Mark III database is not truly uniform; the
effort to unify the five individual sections of Mark III to a
common linewidth, magnitude system is somehow deficient.}  The
Mark III sample is a collection of available data and samples
different directions far from uniformly.  There is a large
concentration of galaxies toward Perseus-Pisces contributed by
Willick (1990), and a very large concentration of galaxies in the
Southern sky contributed by Mathewson \etal (1992) on the
opposite side of the sky.  More importantly, different techniques
are used in differet directions.  Any inconsistencies within the
Mark III catalog are likely to translate into dipole effects, and
indeed the dominant residuals between Mark III and \iras is a
dipole aligned along the center of the MAT subsample.

\item{ } Willick \etal (1996a,b) describe the comparison of the
different pieces of the Mark III sample.  It is obviously
critically important to insure that the relative photometry and
linewidth measurements are standardized, and it is possible that
the Mark III data has zero point calibration problems between its
different subpieces.  As described above, we have tested the
sensitivity of the ITF field to the known statistical uncertainty
of the zero-point offsets, finding a negligible effect.  If
calibration mismatches are the source of the problem, then the
error must be considerably in excess of the measured statistical
errors.  It would be especially valuable to obtain additional
data to better tie together the different pieces of Mark III, and
to improve the sky coverage in the North and away from the dense
cluster centers.  The Mark III sample is vulnerable to any
systematic nonuniformities between Han-Mould Northern versus
Southern clusters, because this subsample is used to
renormalize the different pieces of Mark III to one system.
  A new survey of Tully-Fisher distances for
300 galaxies in the zone $5000<cz<7000 \kms$ around the entire
sky is just now getting under way (Strauss \etal 1996), with the
goal of better calibration of the whole-sky samples.  

\item{ } A considerable portion of the Mark III sample is derived from
relatively low galactic latitude data, where corrections for
galactic extinction are significant.  Recently Schlegel,
Finkbeiner, and Davis (1996) have been studying whether a new
extinction map, based on the COBE/DIRBE data, leads to
significantly different flow fields.  The question is still under
investigation, but the answer appears to be no.

\item{(5.)} {\it The Tully-Fisher calibration is  dependent on environment and
the residual field is caused by changes in the TF parameters for different
sections of the sky}.  Although it is impossible to prove that the TF
relation is independent of environment, it is not easy to imagine
schemes whereby the relation is coherently different in one portion of
the sky compared to another.  It is much easier to imagine environmental
dependence leading to larger scatter of the TF relation.  It is important
to note that the most serious zone of discrepancy, in the Centaurus region,
has been extensively studied with the $D_n - \sigma$ relation for ellipticals 
(Lynden-Bell \etal 1988) as well as recent surface brightness fluctuation 
measurements using ellipticals (Dressler 1994; Tonry and Dressler 1995).  All 
three 
distance indicators suggest strong outflow in the Centaurus region, which
the \iras maps fail to predict.

Improving the \iras maps by going to denser
sampling, either using the PSC catalog (Saunders \etal 1995), or by 
supplementing the \iras catalog with optical data in regions not too badly
extincted (e.g. Santiago \etal 1995), 
will be an important next step to attempt to resolve the 
discrepancies here presented.  
Based on previous analyses (Hudson 1994; Freudling \& da Costa
1994), these improvements in
the gravity map are unlikely to resolve the problem.  
More uniform full-sky Tully-Fisher catalogs, where the line widths and
magnitudes are estimated in a consistent fashion for the entire sample,
will be an enormous step forward, but this is a major undertaking.

\bigskip
{\cl {\bf 7. Summary}}

We have presented a method for comparing the velocity field derived from 
Tully-Fisher type data with the gravity field derived from full-sky 
redshift surveys of galaxies, such as the \iras surveys.  The chief strength of 
the method is that the two fields are filtered through the same set of 
low resolution modes, so that one is assured they indeed have the same 
resolution. In spite of working in the `inverse' direction, we are able to
produce pictures of the measured fields.

Although the method works extremely well in mock catalogs, with modal
coefficients 
fully consistent with the expected noise, we find a much poorer
agreement for the real data.  
This discrepancy is not consistent
with random errors, and 
 may be indicative of a systematic error in one or both
of the datasets.  We derive a ``best" value of $\beta = 0.4-0.6$, 
in reasonable agreement with the value derived from 
the \velmod analysis, though considerably
smaller than the values ($\beta\simeq 1.0$) typically obtained
from applications of the POTENT algorithm. Because the
$\chi^2$ for the fit of the ITF velocity field to
the \iras gravity field is 100 for 55 degrees of freedom, the
fields cannot be said to agree in detail. 
We therefore urge extreme caution in 
concluding that $\beta$ has been measured in large scale flows.  
It is worth emphasizing that there is qualitative
agreement between the velocity and gravity fields, particularly
at distances $\simlt 3000 \kms.$ However, until the
discrepancies on larger scales can be resolved,
and all methods give consistent answers, 
we must prudently consider the value of $\beta$ to be
an open question.

Future work should greatly improve this situation, as more data, and more
precise data, become available. 
The Mark III catalog is 
 not optimal for the analysis presented here,
since the sky coverage is  relatively nonuniform. 
For a given
allocation of telescope time, the optimal strategy is to sample the sky
to a given depth as uniformly as possible.  
Full sky samples are difficult to calibrate
in a uniform manner, but are worth the effort, since they provide the
only route to estimation of the low order multipoles of the local velocity
field.

\vfill\eject
{\bf Acknowledgments}

This work was supported in part by NSF grant AST 92-21540 and NASA grant
NAG 51360. 
AN acknowledges the support of a PPARC postdoctoral fellowship.
We thank Simon White for the use of the large $N$-body simulation and
we acknowledge helpful discussions with Saleem Zaroubi.
JAW would like to acknowledge the contributions of
his collaborators on the Mark III project,
Avishai Dekel, Sandra Faber, David Burstein, and St\'ephane
Courteau. We thank Michael Strauss and the referee Michael
Vogeley for numerous helpful comments that have substantially 
improved the manuscript.

\bigskip\bigskip
\centerline{\bf Table 1: Coefficient Comparison}
$$\vbox{ \settabs 2 \columns
\+ $\beta$ &  $\chi^2$ \cr
\+ \cr
\hrule
\+ \cr
\+ 0.1& 155 \cr
\+ 0.2& 124 \cr
\+ 0.3& 107.5 \cr
\+ 0.4& 100.5 \cr
\+ 0.5& 100 \cr
\+ 0.6& 102.8 \cr
\+ 0.7& 112 \cr
\+ 0.8& 121 \cr
\+ 0.9& 134 \cr
\+ 1.0& 147 \cr

}$$

\vfill\eject
\bigskip
\cl{\bf REFERENCES}
\bigskip
\def\ref{\par\noindent\hangindent 15pt}
\def\apj#1{{\it Astrophys. J.} { #1}}
\def\apjl#1{{\it Astrophys. J. (Lett.)} { #1}}
\def\apjs#1{{\it Astrophys. J. (suppl.)} { #1}}
\def\mn#1{{\it M.N.R.A.S.} { #1}}

\ref Aaronson, M. Huchra, J., Mould, J., Schechter, P., \& Tully, R. B. 1982,
\apj{258}, 64

\ref Bertschinger, E., Dekel, A., Faber, S. M., Dressler, A., \& Burstein, D.
1990 \apj{364}, 370

\ref Bond, J. R. 1995, Phy. Rev. Let., 74, 4369

\ref Bucher, M. \& Turok, N. 1995 preprint, hep-ph/9503393

\ref Bunn, T. 1995, Ph.D thesis, UC Berkeley

\ref Connolly, A. J., Szalay, A. S., \& Bershady, M. A., Kinney, A. L. 
\etal 1995, Astron. J., 110, 1071

\ref Davis, M., Strauss, M., \& Yahil, A. 1991 \apj{372}, 394

\ref Davis, M., \& Nusser, A. 1995, in {\it The Maryland Meeting on Dark
 Matter}, AIP Conference Proceedings 336, ed S. Holt \& C. Bennett, p. 361

\ref Davis, M., Tonry, J., Huchra, J., \& Latham, D. 1980, \apjl{238}, L113

\ref Dekel, A. 1994, Ann. Rev. of A \& A, {\bf 32}, 371

\ref Dekel, A. 1995, private comm.

\ref Dekel, A., Bertschinger, E., \& Faber, S. M. 1990, \apj{364}, 349

\ref Dekel, A., Bertschinger, E., Yahil, A., Strauss, M., Davis, M., \&
Huchra, J. 1993, \apj{412}, 1

\ref Dressler, A. 1994, in {\it Cosmic Velocity Fields}, ed. F.R. Bouchet 
\& M. Lachi\`eze-Rey, Editions Fronti\`eres, p. 9

\ref Faber, S. \& Burstein, D. 1988 in {\it  Large Scale Motions in the 
Universe}, Princeton Univ. Press, ed. V Rubin \& G Coyne, p. 115 

\ref Fisher, K.B., Davis, M., Strauss, M.A., Yahil, A., \& Huchra, J.P. 1994a, 
\mn{267}, 927.

\ref Fisher, K.B., Huchra, J., Strauss, M.A., Davis, M. , Yahil, A., \&
Schlegel, D. 1995a, \apjs{100}, 69 

\ref Fisher, K.B., Lahav, O., Hoffman, Y., Lynden-Bell, D.,  \&
Zaroubi, S. 1995b, \mn{272}, 885

\ref Fisher, K.B.,  Scharf, C.A., \& Lahav, O., 1994b, \mn{266}, 219

\ref Freudling, W., \& da Costa, L., in {\it Cosmic Velocity Fields}, 
ed. F.R. Bouchet \& M. Lachi\`eze-Rey, Editions Fronti\`eres, p. 187.

\ref Gorski, K., Davis, M., Strauss, M., White, S., \& Yahil, A. 1989 
\apj{344}, 1

\ref Gorski, K. 1994,  ApJL. , 430,  L85

\ref Hudson, M. 1994, \mn{266}, 475

\ref Hudson, M.J., Dekel, A., Courteau, S., Faber, S.M.,
\& Willick, J.A. 1995, \mn{274}, 305

\ref Jackson, D. 1962, {\it Classical Electrodynamics} Wiley

\ref Jacoby, G. H., Branch, D., Ciardullo, R., Davies, R. L., \etal 1992,
{\it Pub. A. S. Pacific}, { 104}, 599


\ref Kaiser, N. 1987, \mn{227}, 1

\ref Kauffman, G., Nusser, A., \& Steinmetz, M. 1996, \mn{} in press. 

\ref Kolatt T., Dekel, A., Ganon, G, \& Willick J 1996, \apj{458}, 419

\ref Kraan-Korteweg, R.C., Woudt, P.A., Cayatte, V., Fairall, A.P.,
Balkowski, C., \& Henning, P.A.  1996, Nature, 379, 519

\ref Kraan-Korteweg, R.C., Loan, A. J., Burton, W.B., Lahav, O. \etal 1994,
Nature, 372, 77

\ref Lahav, O. 1991 in {\it Highlights of Astronomy}, vol.9, ed. Bergeron, J.,
Kluwer, Dordrecht

\ref Liddle, A.R. \& Lyth, D.H. 1993, Physics Reports, 231, 1

\ref Lynden-Bell, D. 1991, in {\it  Statistical Challenges in Modern
Cosmology}, eds. Babu, G.B \& Feigelson, E.D.

\ref Lynden-Bell, D., Faber, S. M., Burstein, D., Davies, R. L., Dressler, A.,
Terlevich, R. J., \& Wegner, G. 1988, \apj{326}, 19

\ref Miller, A., Davis, M., \& White, S.D.M. 1996, preprint

\ref Nusser, A., \& Davis, M. 1994, \apjl{421}, L1

\ref Nusser, A., \& Davis, M. 1995, \mn{276}, 1391

\ref Nusser, A., Dekel, A., Bertschinger, E., \& Blumenthal, G. R. 1991,
     \apj{379}, 6

\ref Ostriker, J, Peebles, P.J.E, \& Yahil, A. 1974, \apjl{193}, L1

\ref Peebles, P. J. E. 1980, {\it The Large Scale Structure of the Universe},
Princeton Press, p. 65

\ref Peebles, P. J. E. 1992, in {\it Relativistic Astrophysics and Particle
Cosmology}, Texas/PASCOS 92 Symp. ed C. W. Akerlof \& M. A. Srednicki (Ann. NY
Acad Sci. vol 688), 84

 \ref Press, W.H., Teukolsky, S.A., Vetterling, W.T., \& Flannery, B.P. 1992,
{\it Numerical Recipes}, (Second Edition) 
Cambridge University Press

\ref Reg\"os, E., \& Szalay, A. 1989, \apj{345}, 627

\ref Roth, J. 1993, in {\it Cosmic Velocity Fields}, ed. F. Bouchet \&
M. Lachieze-Rey, Editions Frontieres, p233.

\ref Schechter, P. 1980 \apj{85}, 801

\ref Schlegel, D., Davis, M., Summers, F., \& Holtzman, J. A. 1994, \apj{427}
527

\ref Schlegel, D, Finkbeiner, D, \& Davis, M. 1996, in preparation


\ref Strauss, M. A., \& Davis, M. 1988, in {\it Large Scale Motions in the 
Universe}, ed. V Rubin \& G Coyne, Princeton Univ. Press, p. 256

\ref Strauss, M.A., Davis, M., Yahil, A., \& Huchra, J. 1992, \apj{385}, 421

\ref Strauss, M. A., \& Willick, J. 1995, Physics Reports, 261, 271

\ref Strauss, M.A., Willick, J.A., Schlegel, D., Postman, M.,
\& Courteau, S. 1996 in preparation

\ref Strauss, M.A., Yahil, A., Davis, M., Huchra, J.P., \&
Fisher, K. 1992 \apj{397}, 395 

\ref Therrien, C. W. 1992, {\it Discrete Random Signal and Statistical 
Signal Processing}, (New Jersey: Prentice-Hall)

\ref Tonry, J, \& Davis, M. 1981 \apj{246}, 680 

\ref Tonry, J, \& Dressler, A. 1995, in preparation

\ref Vogeley, M. S. , \& Szalay, A. S. 1996, preprint astro-ph/9601185

\ref Willick, J.A. 1990, \apjl{351}, L5

\ref Willick, J.A. 1994, \apjs{92}, 1.

\ref Willick, J.A., Courteau, S, Faber, S., Burstein, D., \etal 1995, 
\apj{446}, 12 

\ref Willick, J.A., Courteau, S, Faber, S., Burstein, D., 
Dekel, A., \& Kolatt, T. 1996a, \apj{457}, 460 

\ref Willick, J.A. \etal 1996b, in preparation

\ref Willick, J.A. \etal 1996c, in preparation

\ref Yahil, A. 1988, in  {\it Large Scale Motions in the 
Universe}, Princeton Univ. Press, ed. V Rubin \& G Coyne, p. 219

\ref Yahil, A., Strauss, M. A., Davis, M., \& Huchra, J. 1991 \apj{372}, 380

\vfill\eject
\bigskip

\myfig {1} {20.0}
{\includegraphics{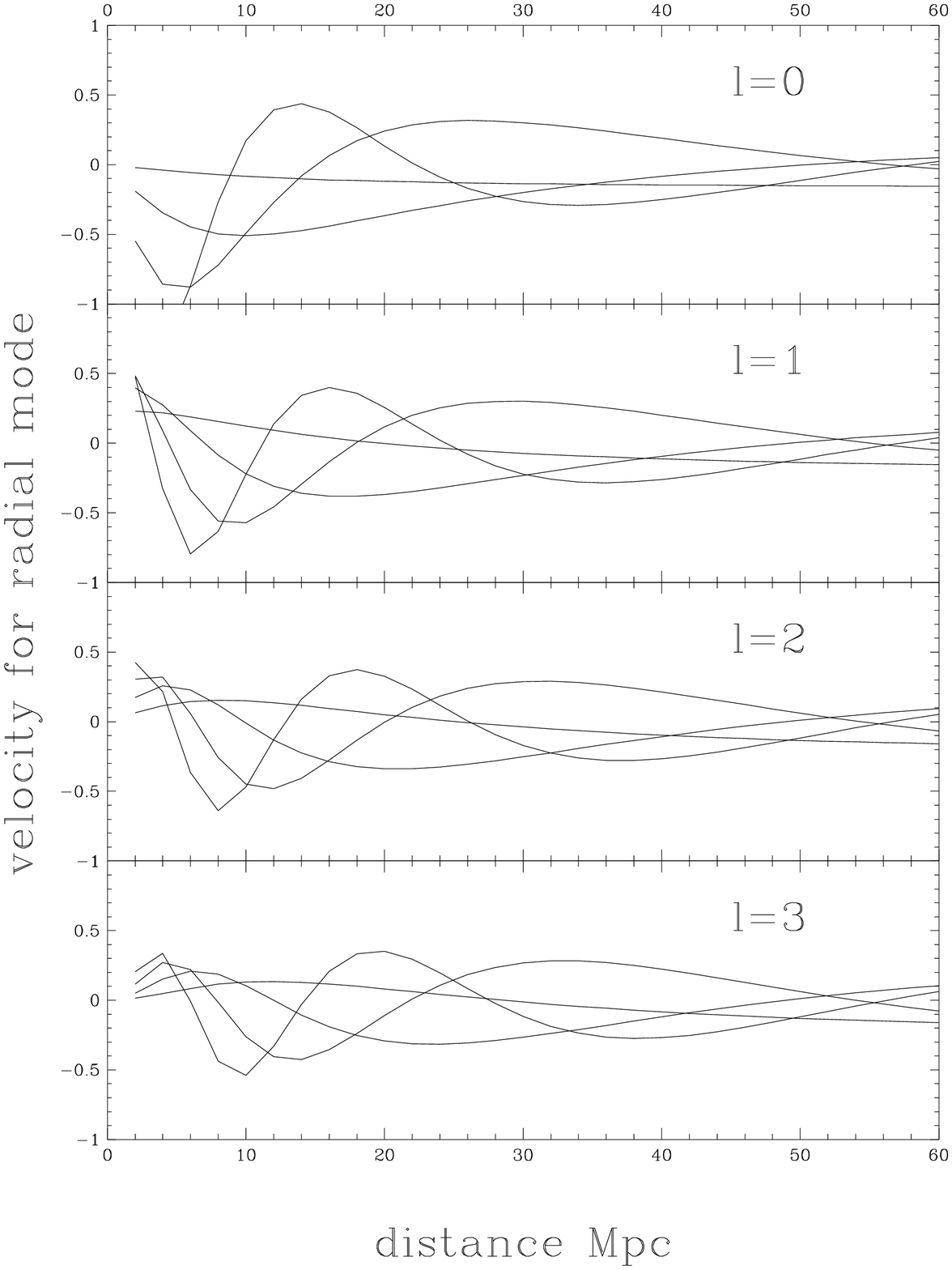}}
{The amplitude of the radial variation of the mode velocities 
 versus cz for l=0-3. Note that the radial resolution drops sharply with
 redshift.}

\np
\myfig {2} {20.0}
{\vskip -3. truein
{\it this figure available by anonymous ftp at 
magicbean.berkeley.edu/pub/marc/m3itf/ }
\vskip 3. truein}
{ The $u_{itf}$ sky projection as seen in the LG frame for
the mock catalog, in galactic
coordinates.  The open circles are points that are flowing inward
and the stars are points flowing outward.  The size of the symbols is 
proportional to the flow velocity, with a key
showing $500 \kms$ flow.  Note the three separate panels of
redshift interval.  Note also the dominance of the dipole mode, especially for
the outer shell.  This is the signature of the reflex motion of the LG.}

\np
\myfig {3} {20.0}
 {\vskip -3. truein
{\it this figure available by anonymous ftp at 
magicbean.berkeley.edu/pub/marc/m3itf/ }
\vskip 3. truein}
{The $u_{itf}$  field for points within $30^\circ$ of  the 
Supergalactic Plane, plotted in this plane, for the
mock catalog, in the LG frame.  Solid symbols and solid lines are outflowing 
``galaxies" with
the symbol drawn at the distance and the line drawn to the measured
LG redshift.  Open symbols and dashed lines correspond to 
``galaxies" infalling to the LG.  Note again the strong dipole pattern of
the flow at $z> 3000 \kms.$ }

\np
\myfig {4} {20.0}
 {\vskip -3. truein
{\it this figure available by anonymous ftp at 
magicbean.berkeley.edu/pub/marc/m3itf/ }
\vskip 3. truein}
{ (a) The true, unfiltered peculiar velocities, 
$u_{true}$ of the points in the
mock catalog, plotted in the Supergalactic Plane.  
(b:) The true peculiar velocities, $u^f_{true}$ of the mock catalog
points, after filtering by the 56 mode expansion.  Note how the
large scale behavior of panel a is reproduced in panel b.  The high
spatial frequency signal, which is very difficult to reproduce in the
\iras\ gravity field, is effectively eliminated in Figure 4b.
(c:) The \iras\ predicted velocity field for the
mock catalog,  $u_{iras}$, plotted in the Supergalactic Plane.
(d:) The same as panel (c), but now filtered by the
56 mode expansion, $u^f_{iras}$.}

\np
\myfig {5} {20.0}
{\includegraphics{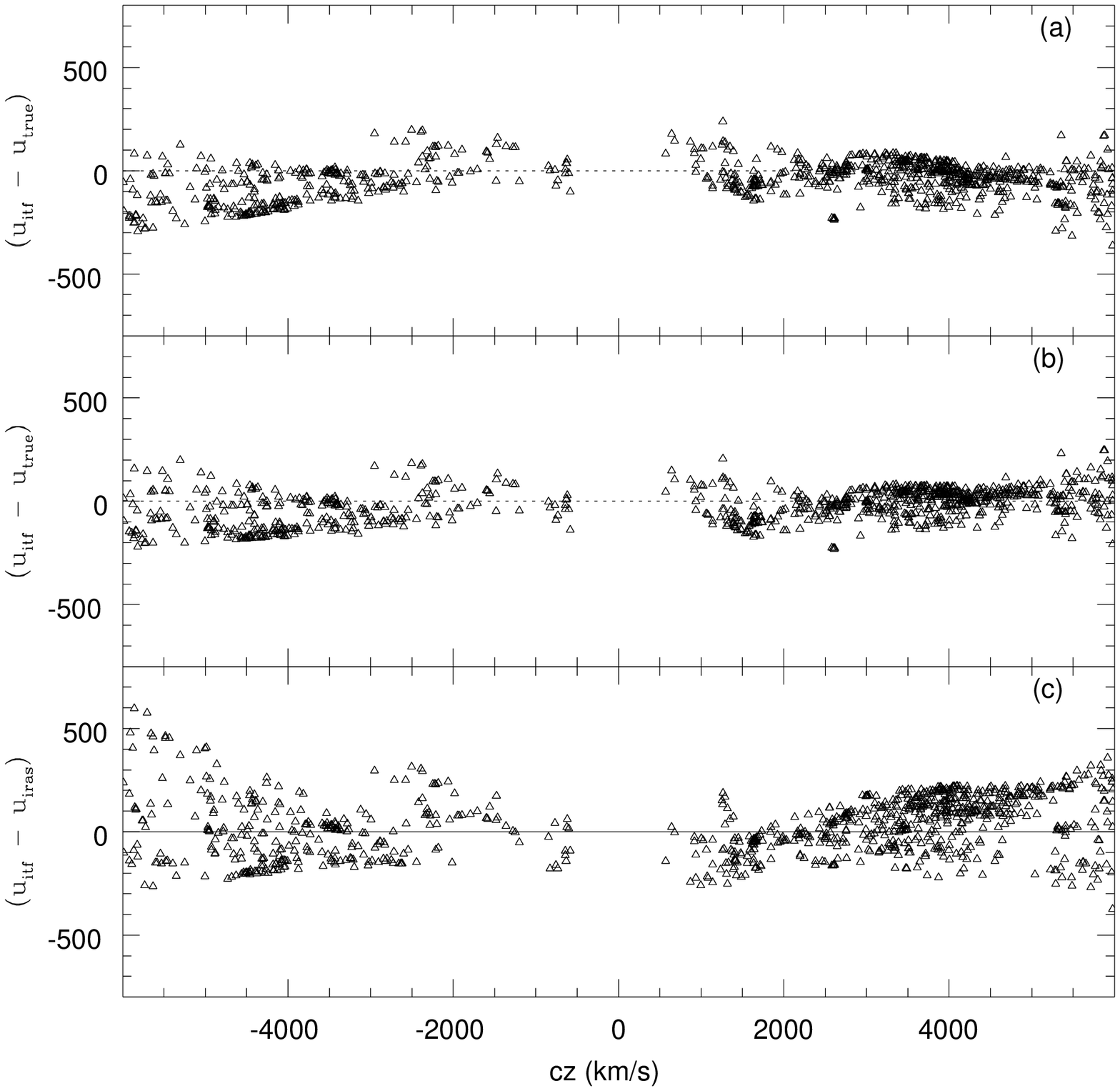}}
{(a:) The residuals $u_{itf} - u^f_{true}$ as a function
of cz of points 
in the mock catalog within
$60^\circ$ of the apex and anti-apex direction (l,b) = (270,30), before
correction for the bias in the slope parameter $\gamma$.  
(b:)  The same plot $u_{itf} - u^f_{true}$ after the bias in the 
slope parameter is removed. (c:) The residuals of 
$u_{itf} - u^f_{iras}$ for the same galaxies. }

\np
\myfig {6a} {20.}
 {\vskip -3. truein
{\it this figure available by anonymous ftp at 
magicbean.berkeley.edu/pub/marc/m3itf/ }
\vskip 3. truein}
{Residuals $u_{itf}-u^f_{true}$,  
in sky slices for  mock catalog, using the same
notation as in Figure 2. 
}

\np

\myfig {6b} {20.0}
 {\vskip -3. truein
{\it this figure available by anonymous ftp at 
magicbean.berkeley.edu/pub/marc/m3itf/ }
\vskip 3. truein}
{ Residuals $u_{itf}-u^f_{iras}$ in the
same projection. Note the small amplitude of these residuals
versus Figure 2, and the absence of significant dipole contribution to
these residuals. Note that the amplitude of the residuals in (6b) are somewhat
larger than those of (6a), because of the \iras contributed noise.}

\np
\myfig {7} {20.0}
{\includegraphics{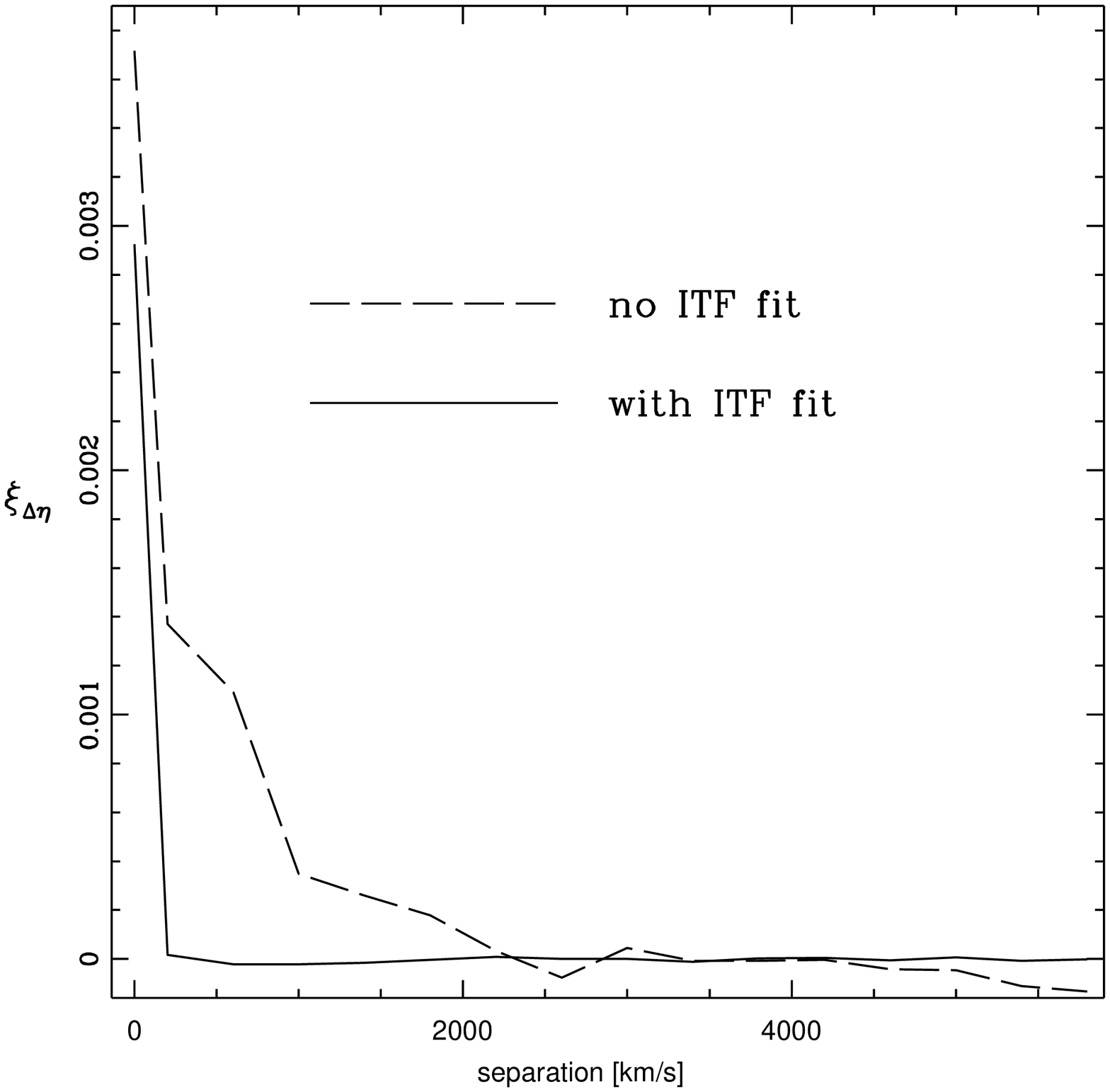}}
{The auto-correlation function of the $\eta$ residuals of the
Mark III galaxies versus redshift space separation.  
The dashed curve results when the 56 mode coefficients are set to 0, while
the solid curve is after the 56 modes are fit to the Mark III data. }

\np
\myfig {8} {20.0}
 {\vskip -3. truein
{\it this figure available by anonymous ftp at 
magicbean.berkeley.edu/pub/marc/m3itf/ }
\vskip 3. truein}
{The ITF velocity field $u_{itf}$ for the Mark III
galaxies using the same sky projection as Figure 2.}

\np
\myfig {9} {20.0}
 {\vskip -3. truein
{\it this figure available by anonymous ftp at 
magicbean.berkeley.edu/pub/marc/m3itf/ }
\vskip 3. truein}
{The ITF velocity field $u_{itf}$: for Mark III
galaxies within $30^\circ$ of the supergalactic plane, using the
same notation as in Figure 3.}

\np
\myfig {10} {20.0}
 {\vskip -3. truein
{\it this figure available by anonymous ftp at 
magicbean.berkeley.edu/pub/marc/m3itf/ }
\vskip 3. truein}
{The sky projection of the  \iras predicted flow field $u_{iras}$ for 
$\beta=0.4$ for the Mark III galaxies.}

\np
\myfig {11} {20.}
 {\vskip -3. truein
{\it this figure available by anonymous ftp at 
magicbean.berkeley.edu/pub/marc/m3itf/ }
\vskip 3. truein}
{The supergalactic plane projection for the \iras predicted
flow field $u_{iras}$ for $\beta=0.4$ for the  Mark III galaxies.}

\np
\myfig {12} {20.0}
 {\vskip -3. truein
{\it this figure available by anonymous ftp at 
magicbean.berkeley.edu/pub/marc/m3itf/ }
\vskip 3. truein}
{The sky projection of the residuals  $u_{itf} - u_{iras}$ for
 $\beta=0.4$.  This is simply the difference between Figures 8 and 10.
Compare this field to Figure 6b. }

\np
\myfig {13} {20.0}
 {\vskip -3. truein
{\it this figure available by anonymous ftp at 
magicbean.berkeley.edu/pub/marc/m3itf/ }
\vskip 3. truein}
{The same as Figure 12, now for $\beta=0.7$.}

\np
\myfig {14} {20.}
{\includegraphics{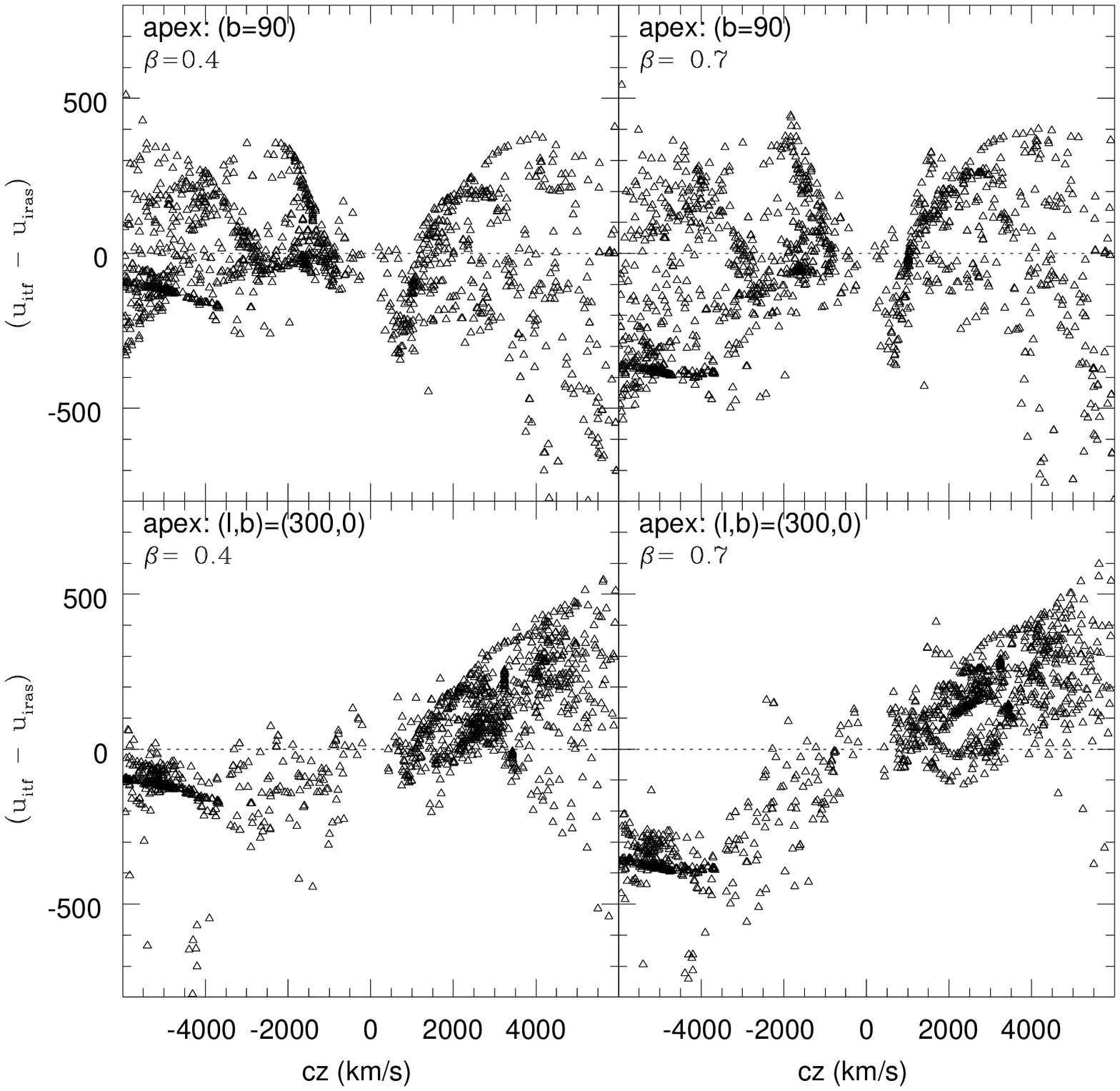}}
{(left): The residual $u_{itf} - u_{iras}$ of all points within 60$^\circ$ 
of the apex and antiapex of (l,b) = (0,90) and (300,0), for $\beta = 0.4$.
(right) The same residual for the case $\beta=0.7$.  Note the linearly
growing dipole residual toward (300,0).}

\np
\myfig {15} {20.0}
{\includegraphics{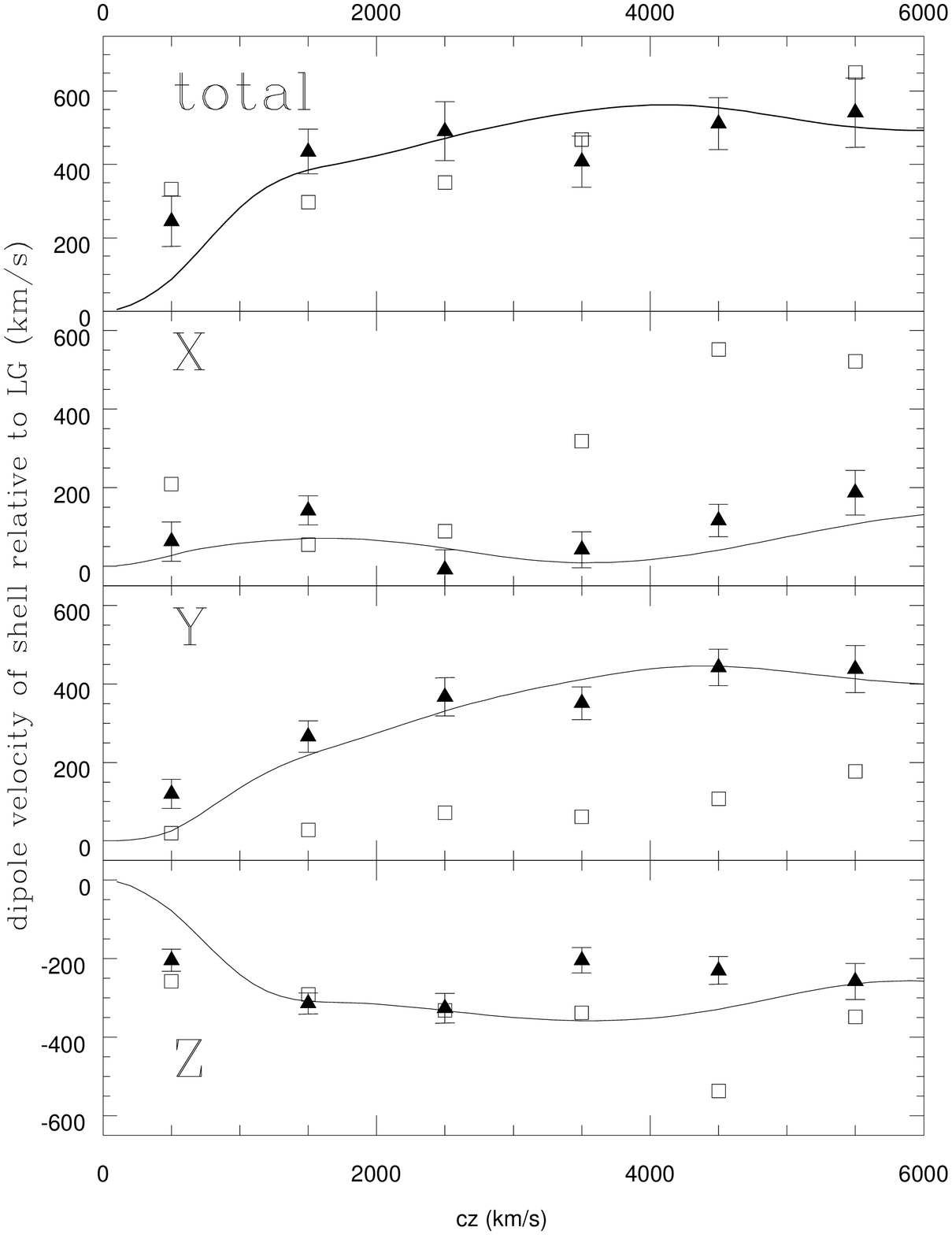}}
{The dipole field derived from the Mark III galaxies. The ITF 
dipole is plotted as the open symbols, while the \iras dipole for $\beta = 0.6$
is plotted as the closed symbols. 
The X,Y,Z panels show the three Galactic components 
of the dipole and the top panel is their quadrature sum.  The
solid line is the full \iras dipole field calculated from Equation 4.  The 
Mark III galaxies are binned in redshift shells of $1000 \kms$ thickness.
Note that the \iras inferred dipole for the Mark III symbols tracks the solid 
line, while the X and Y 
components of the ITF dipole are quite deviant from the line. }

\np
\myfig {16} {20.0}
{\includegraphics{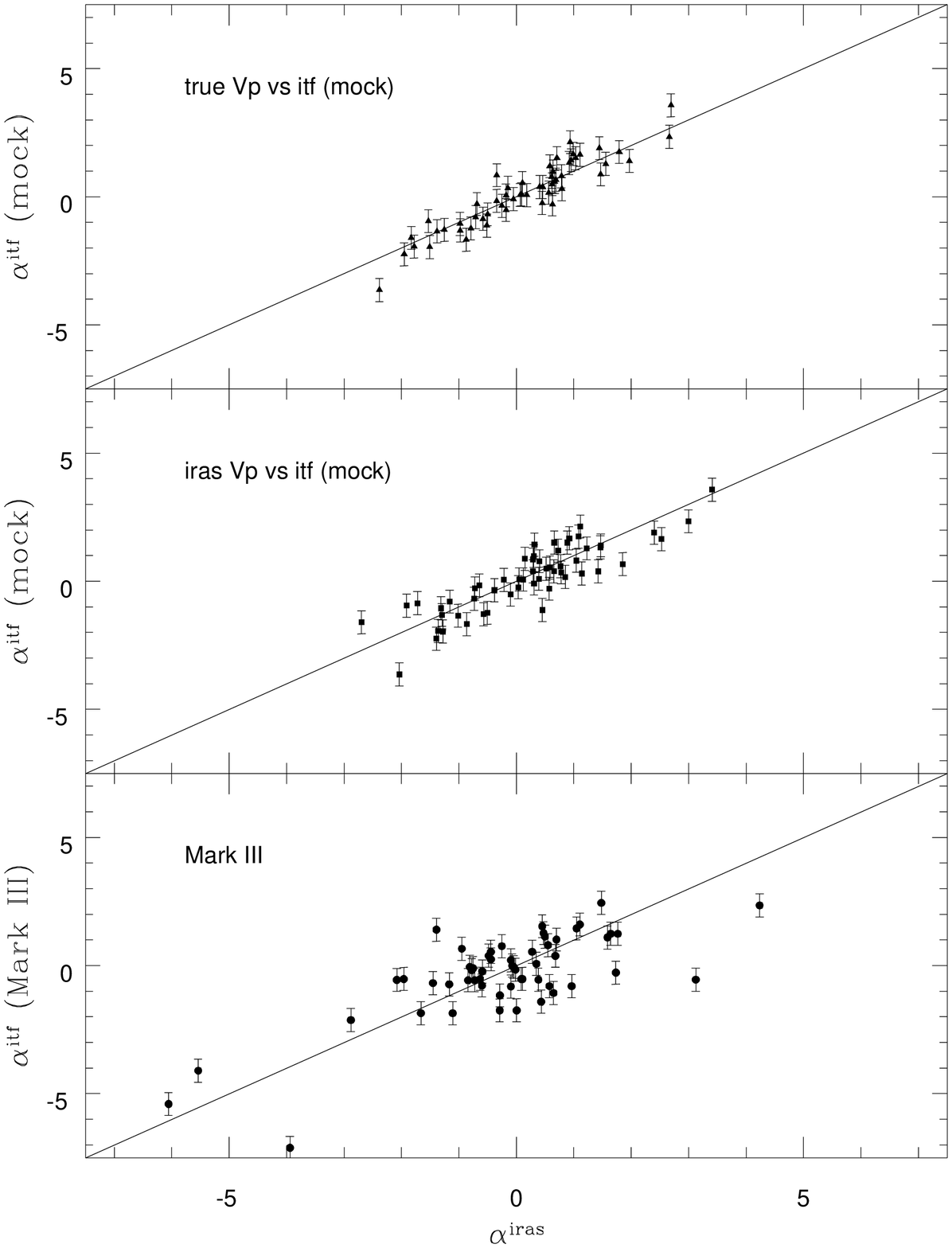}}
{(top:) A scatter diagram of $\alpha^{itf}$ versus the
true coefficients $\alpha^{true}$ 
for the mock catalog.  The ITF errors  are shown.
(middle:) Scatter of $\alpha^{itf}$ versus $\alpha^{iras}$ 
for the mock catalog. The increased scatter is  induced by the
errors in $\alpha^{iras}$, not shown because of the covariance.
(bottom:) The scatter of $\alpha^{itf}$ versus 
$\alpha^{iras}$  for the
real Mark III data ($\beta=0.6$). Note the qualitative agreement of
the mode amplitudes but the further increase in scatter. }

\bye